\documentclass[conference]{IEEEtran}
\IEEEoverridecommandlockouts

\usepackage{cite}
\usepackage{amsmath,amssymb,amsfonts}
\usepackage{algorithmic}
\usepackage{graphicx}
\usepackage{textcomp}

\usepackage{comment}
\usepackage[capitalize]{cleveref}
\usepackage[table,xcdraw]{xcolor}
\usepackage{hhline}
\usepackage{multirow}
\usepackage{braket}
\usepackage{pifont}
\usepackage{url}

\newcommand{\figsize}{0.98\linewidth}
\newcommand{\archname}{D3-ROT}

\makeatletter
\newcommand{\linebreakand}{%
  \end{@IEEEauthorhalign}
  \hfill\mbox{}\par
  \mbox{}\hfill\begin{@IEEEauthorhalign}
}
\makeatother

\def\BibTeX{{\rm B\kern-.05em{\sc i\kern-.025em b}\kern-.08em
    T\kern-.1667em\lower.7ex\hbox{E}\kern-.125emX}}
\begin{document}

\title{NAQsim: Full-Stack Architecture Simulation Framework for Fast and Space-Efficient\\Neutral Atom Quantum Computing\\

\thanks{\IEEEauthorrefmark{1}Corresponding author.\\}

\thanks{
This work was supported in part by JSPS KAKENHI Grant Numbers JP24K02915, JP25K21175, JP25K21176, JP25K24539, and JP25K03094;
JST Moonshot R\&D Program Grant Numbers JPMJMS256L and JPMJMS256E; 
MEXT Q-LEAP Grant Numbers JPMXS0120319794 and JPMXS0118068682; 
JST CREST Grant Numbers JPMJCR23I4, JPMJCR24I4, and JPMJCR25I4;
and the Feasibility Study on the future HPCI.
}

}

\author{
% First line
\IEEEauthorblockN{Yosuke Ueno}
\IEEEauthorblockA{
\textit{Nanofiber Quantum Technologies}\\
Tokyo, Japan\\
\textit{RIKEN Center for Quantum Computing}\\
Wako, Japan\\
yosuke.ueno@riken.jp
}
\and
\IEEEauthorblockN{Shinichi Sunami}
\IEEEauthorblockA{
\textit{Nanofiber Quantum Technologies}\\
Tokyo, Japan\\
\textit{University of Oxford}\\
Oxford, United Kingdom\\
shinichi.sunami@nano-qt.com
}
\and
\IEEEauthorblockN{Toshihide Hinokuma}
\IEEEauthorblockA{
\textit{Nanofiber Quantum Technologies}\\
Tokyo, Japan\\
toshihide.hinokuma@nano-qt.com
}
\and

\linebreakand

% Second line
\IEEEauthorblockN{Yasunari Suzuki}
\IEEEauthorblockA{
\textit{RIKEN Center for Quantum Computing}\\
Wako, Japan\\
yasunari.suzuki@riken.jp
}
\and
\IEEEauthorblockN{Akihisa Goban}
\IEEEauthorblockA{
\textit{Nanofiber Quantum Technologies}\\
Tokyo, Japan\\
akihisa.goban@nano-qt.com
}
\and
\IEEEauthorblockN{Hayata Yamasaki}
\IEEEauthorblockA{
\textit{Nanofiber Quantum Technologies}\\
Tokyo, Japan\\
\textit{The University of Tokyo}\\
Tokyo, Japan\\
hayata.yamasaki@gmail.com
}

\linebreakand

% Third line
\IEEEauthorblockN{Teruo Tanimoto}
\IEEEauthorblockA{
\textit{Kyushu University}\\
Fukuoka, Japan\\
tteruo@kyudai.jp
}
\and
\IEEEauthorblockN{Ilkwon Byun\IEEEauthorrefmark{1}
}
\IEEEauthorblockA{
\textit{Kyushu University}\\
Fukuoka, Japan\\
ilkwon@kyudai.jp
}
}

\maketitle
\thispagestyle{plain}
\pagestyle{plain}

%%%%%% -- PAPER CONTENT STARTS-- %%%%%%%%
\begin{abstract}
Technological advances in neutral-atom platforms have opened a new path toward designing efficient protocols for fault-tolerant quantum computing (FTQC).
However, each physical operation is still orders of magnitude slower than on other platforms, such as superconducting qubits. 
Therefore, architects must identify fast and efficient FTQC architectures, which require reliable modeling tools to explore various design choices across the software, classical hardware, and quantum device stacks of neutral-atom platforms.

In this paper, we propose NAQsim, an open-source simulation framework for neutral-atom FTQC architectures based on transversal gates.
As its key feature, NAQsim enables detailed full-stack architecture evaluation that opens opportunities to explore previously overlooked performance bottlenecks.
As a first use case for NAQsim, we identify one such bottleneck, patch rotations, perform full-stack co-optimization, and finally derive a near-rotation-free architecture (D3-ROT). 
For practical FTQC benchmark workloads, D3-ROT achieves a 2.27$\times$ speedup from this single bottleneck alone, with minimal footprint overhead. 
These results point to a much broader space of full-stack optimizations that NAQsim makes accessible for transversal surface-code architectures and beyond. 
\end{abstract}

\section{Introduction}
The paradigm of neutral-atom (NA) quantum computing has rapidly reshaped both experiment and theory of fault-tolerant quantum computing (FTQC).
For example, the development of fast in-place two-qubit gates~\cite{Radnaev2025} and advanced readout techniques~\cite{lis2023midcircuit,Radnaev2025,anand2024dual-species} has enabled fast shuttling-free error syndrome measurement~(ESM)~\cite{sunami2025transversalsurfacecodegamepowered}. 
In parallel, theoretical breakthroughs, such as the introduction of transversal fault tolerance~\cite{zhou2024algorithmic}, have drastically reduced the number of required ESM rounds to $O(1)$ per logical operation.

Despite remarkable progress, NA quantum computers remain slow due to performance bottlenecks in implementing logical operations. 
Even though the ESM latency has been greatly reduced, logical operations (e.g., transversal CX and H) are still prohibitively slow as they heavily rely on acousto-optic deflector (AOD)-based atom shuttling. 
Specifically, the AOD-based logical operations can take a few hundred microseconds to several milliseconds, potentially much longer than a shuttling-free ESM~\cite{sunami2025transversalsurfacecodegamepowered}.
Due to the logical-operation bottleneck, state-of-the-art neutral-atom architectures still suffer from a huge performance gap relative to superconducting qubit platforms (58$\times$ in our analysis).

To close this gap, it is imperative to explore various architectural solutions to reduce logical operation latency. 
However, it is challenging due to the complex interplay among the compiler, control hardware, and qubit plane.
For example, the qubit-plane design limits the available implementations of a logical operation. 
In addition, the implementation of a given instruction in the laser system affects compiler optimizations. 
Most importantly, these tight inter-stack relationships pose significant challenges in addressing the trade-offs among performance, space overhead, and logical error rate~(LER). 

Therefore, architects now need a reliable tool to evaluate architectural ideas holistically. 
However, the community lacks such a comprehensive modeling tool, particularly for NA FTQC architectures.
Although previous studies have proposed modeling tools for superconducting quantum computers~\cite{xqsim, qisim}, they cannot address the unique capabilities (e.g., atom shuttling), control hardware constraints (e.g., AOD route conflict), and qubit plane features (e.g., logical-qubit overlap) of the NA platform.
In addition, existing modeling tools have focused on lattice surgery~\cite{xqsim, qisim, suzuki2026quration}.
However, exploiting neutral atoms' potential requires more efficient FTQC protocols that introduce much greater complexity in modeling.

In this paper, we propose (1) NAQsim, an open-source architecture simulation framework tailored for neutral-atom-based FTQC, (2) novel optimization ideas in compiler, control hardware, and qubit plane stacks, and (3) \archname{}, a near-rotation-free NA architecture driven by our full-stack co-optimization. 
As a first study, we target the transversal surface code, which offers well-defined logical operations and transversal fault tolerance~\cite{zhou2024algorithmic}.

First, we develop NAQsim to comprehensively evaluate the performance, space overhead, and success probability of the input program on the target NA FTQC architecture. 
NAQsim derives its output in three stages: compilation, execution, and estimation. 
The compilation stage generates instruction schedules for the given logical instruction set. 
Next, the execution stage iteratively simulates the scheduled instructions on the target qubit plane, accounting for the control hardware configuration.
Using the execution trace and other inputs, the estimation stage finally outputs execution time, qubit overhead and footprint, and logical error rate.

Next, driven by architecture exploration using NAQsim, we perform full-stack co-optimization to resolve the previously overlooked performance bottleneck, logical-qubit (or patch) rotation. 
Our performance analysis across various practical FTQC workloads shows that patch rotation accounts for 55.3\%--77.6\% of the total execution time. 
To resolve the bottleneck, we propose three architectural solutions: (1) direct rotation with spatial light modulator~(SLM) laser toggling in the control hardware, (2) dedicated rotation patches in the qubit plane, and (3) distribution of rotations in the compiler. 
With these ideas, whose names start with ``D'', we finally propose \archname{}, a near-rotation-free architecture for fast, space-efficient NA quantum computing. 
For our target FTQC workloads, \archname{} achieves an average speedup of $2.27\times$ and up to $2.57\times$, with a modest 22.4\% footprint overhead on average. 

In summary, our work makes the following contributions: 
\begin{itemize}

\item \textbf{Neutral-atom FTQC modeling framework.}
To the best of our knowledge, this is the first full-stack simulation framework for modeling and evaluating neutral-atom FTQC architectures.

\item \textbf{Novel full-stack co-optimization.}
We provide the co-optimization of the compiler, control hardware, and qubit plane for fast and space-efficient neutral-atom FTQC. 

\item \textbf{Open-source tool release.} 
We release our NAQsim framework to the community to help explore future NA FTQC architectures.\footnote{\url{https://github.com/naqsim/naqsim}}

\end{itemize}

\section{Background and Motivation}

%%%%%%%%%%%%%%%%%%%%%%%%%%%%%%%%%%%%%
\subsection{Neutral-atom quantum computer system} \label{sec:atom_qc_overview}

%%%
\subsubsection{Atom trapping and transport}
Neutral atoms are confined at fixed sites in vacuum using optical tweezers generated by a spatial light modulator (SLM)~\cite{kim2016situ} (\cref{fig:atom_qc_overview}(a)).
The SLM traps remain active throughout subsequent operations (\cref{fig:atom_qc_overview}(b)--(f)), serving as the static default positions for atoms.
Neutral-atom systems offer the ability to physically move atoms at runtime using acousto-optic deflectors (AODs)~\cite{bluvstein2022quantum}.
To shuttle an atom, an AOD trap picks it up from an SLM site, moves it by smoothly ramping the AOD drive frequencies, and releases it at the destination SLM trap.
As AOD frequency changes are applied in parallel across all rows (columns), entire rows (columns) of atoms can move simultaneously.
However, this shuttling takes at least several tens of $\mu\mathrm{s}$ even for $\mu\mathrm{m}$-scale distances, making it a dominant time cost for operations with movement.

%%%
\subsubsection{Single-qubit gate}
\begin{figure}
	\centering
	\includegraphics[width=\figsize{}]{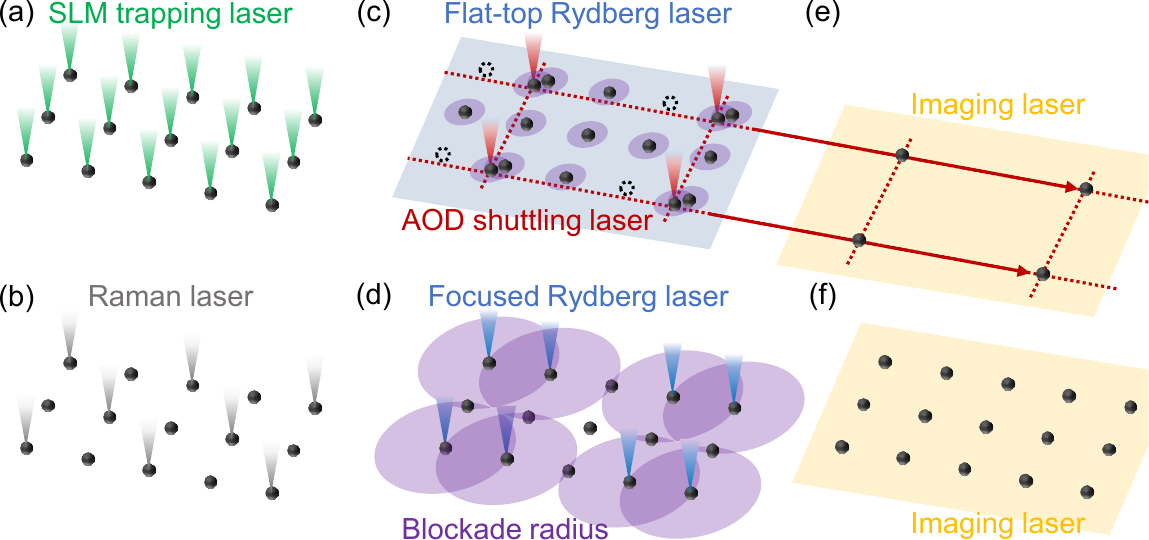}
    \caption{
    (a) Atoms trapped by an SLM laser.
    (b) Selective single-qubit gate via Raman laser.
    (c) Global CZ gate with Rydberg laser. AOD traps bring target pairs within the blockade radius. 
    (d) Selective CZ gate with focused Rydberg beams. 
    (e) Global and (f) selective readout with imaging laser.
    }
	\label{fig:atom_qc_overview}
\end{figure}

Single-qubit gates are performed by driving, for example, Raman transitions between the two qubit states ($\ket{0}$ and $\ket{1}$) using focused laser beams (\cref{fig:atom_qc_overview}(b)).
Gate fidelities exceeding 99.9\% have been demonstrated, with gate times of $100~\mathrm{ns}$~\cite{evered2023high-fidelity,jenkins2022ytterbium}.
We can individually address qubits using Raman beams, thereby selectively applying single-qubit gates to any desired qubits in parallel.

%%%
\subsubsection{Two-qubit gate}
Two-qubit entangling gates utilize the Rydberg blockade~\cite{henriet2020quantum}: a laser excites atoms to a highly excited Rydberg state, and the strong interaction between two Rydberg-excited atoms within a blockade radius $R_\text{blockade}$ enables a controlled-Z (CZ) gate in several hundred nanoseconds~\cite{evered2023high-fidelity,senoo2025}.
Combined with single-qubit H gates, this operation implements CNOT (or CX) gates.
Two implementations are available: global and selective.
In the global approach (\cref{fig:atom_qc_overview}(c)), a broad Rydberg laser illuminates the entire array, and AOD shuttling brings target atoms within $R_\text{blockade}$ of each other while keeping non-targets far apart to avoid undesired interactions~\cite{bluvstein2022quantum}.
In the selective approach (\cref{fig:atom_qc_overview}(d)), tightly focused Rydberg beams address only the target pairs, enabling CZ gates without any shuttling at a switching cost of ${\sim}1~\mu\mathrm{s}$ per addressing pattern change~\cite{Radnaev2025}.

%%%
\subsubsection{Readout}
Qubit readout uses state-dependent fluorescence: an imaging laser excites one qubit state, and the emitted photons are collected by a camera to distinguish $\ket{0}$ from $\ket{1}$ (\cref{fig:atom_qc_overview}(e)--(f)).
As with two-qubit gates, readout has both a global variant with a dedicated imaging zone and an in-place selective variant with negligible crosstalk to neighbors by the use of a hiding beam~\cite{graham2022multi-qubit}, multiple atom species~\cite{anand2024dual-species}, or metastable states of the atom~\cite{lis2023midcircuit}.
The selective approach supports mid-circuit measurement without any atom movement~\cite{lis2023midcircuit, graham2023midcircuit} for in-place ESM~\cite{sunami2025transversalsurfacecodegamepowered}.

%%%%%%%%%%%%%%%%%%%%%%%%%%%%%%%%%%%%%

\subsection{FTQC implementation with neutral atoms}

\begin{figure}
	\centering
	\includegraphics[width=\figsize{}]{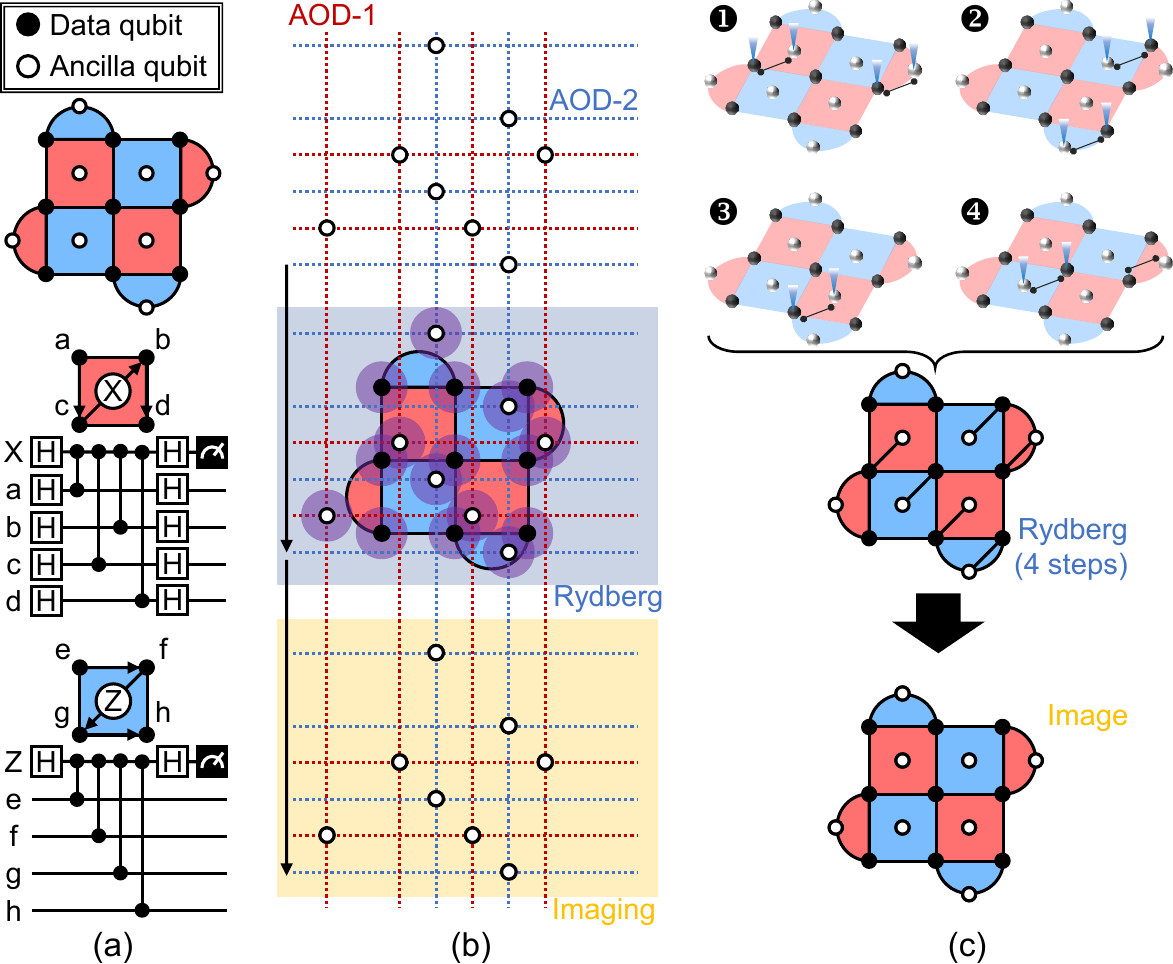}
	\caption{(a) Surface code ($d=3$) and ESM.
    ESM implementation with (b) global and (c) selective Rydberg and imaging lasers.}
	\label{fig:surface_code_esm}
\end{figure}

%%%
\subsubsection{Quantum error correction (QEC) with the surface code}
The rotated surface code~\cite{bombin2007optimal, fowler2012surface} encodes one logical qubit into $n = d^2$ data qubits on a 2D grid, with $n{-}1$ ancilla qubits measuring X- and Z-type stabilizers (\cref{fig:surface_code_esm}(a)).
ESM is performed by applying a sequence of CZ gates between the neighboring ancilla and data qubits, and measuring the ancilla qubits.
A classical decoder then estimates the errors and determines the corresponding corrections.
When the physical error rate is below the code's threshold of around 1\%, the LER is exponentially suppressed with increasing code distance $d$~\cite{Fowler2012, fowler2012surface}.

%%%%%
\textbf{ESM with global lasers.}
A zoned layout~\cite{bluvstein2024logical} (\cref{fig:surface_code_esm}(b)) uses global Rydberg lasers with AOD-based shuttling.
Data and ancilla qubits must be physically rearranged to bring the desired pairs within $R_\text{blockade}$ for each CZ layer of the ESM, and ancillae are then shuttled to a separate readout zone for global fluorescence imaging.

%%%%%
\textbf{ESM with selective lasers.}
With selective Rydberg gates and selective imaging~\cite{sunami2025transversalsurfacecodegamepowered,Radnaev2025} (\cref{fig:surface_code_esm}(c)), ESM can be executed entirely in place.
Each CZ layer is implemented by sequentially addressing the relevant qubit pairs with focused Rydberg beams (${\sim}1~\mu\mathrm{s}$ per pattern change). 
The ancilla qubits are then measured via selective imaging without disturbing neighboring data qubits.
This shuttling-free approach reduces ESM latency significantly and eliminates the need for separated zones.

\begin{figure}
	\centering
	\includegraphics[width=\figsize{}]{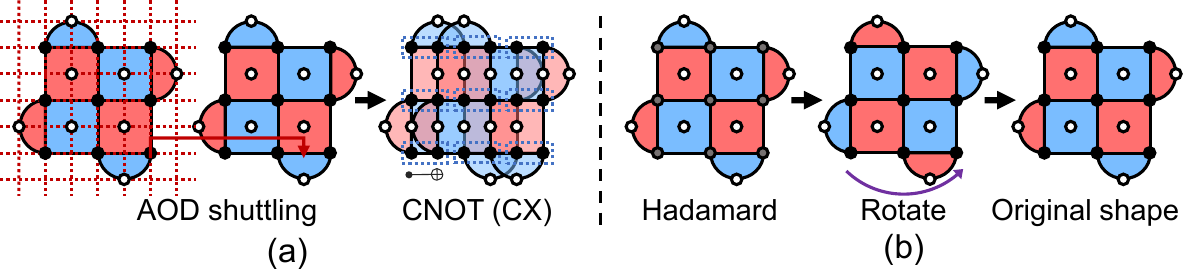}
	\caption{(a) Transversal CX between overlapping logical-qubit patches following movement. 
    (b) Transversal H followed by rotation.}
	\label{fig:transversal_gates}
\end{figure}

%%%
\subsubsection{Transversal logical quantum operations}
For implementing logical operations, lattice surgery~\cite{horsman2012surface, litinski2019game} has been the dominant approach for platforms with fixed 2D nearest-neighbor connectivity, such as superconducting qubits.
In contrast, the shuttling capability of neutral atoms naturally enables transversal logical gates.
We introduce two key logical operations: (1) the transversal CX and (2) the transversal H followed by rotation (Fig.~\ref{fig:transversal_gates}).
The phase (S) gate and the T gate can be implemented by consuming dedicated resource states~\cite{Bravyi2005,sunami2025transversalsurfacecodegamepowered}
using CX gates and logical measurements.

%%%%%
\textbf{Transversal CX gate.}
A logical CX is realized by shuttling one surface-code patch to overlap with another and then applying pairwise physical CX gates between data qubits via selective Rydberg interactions (4 steps, \cref{fig:transversal_gates}(a)).
This requires the SLM lattice spacing to be large enough for an atom array to pass through the gaps between adjacent trap sites during transport.
For the overlapping patches, ESM can be performed independently on each logical qubit using selective lasers, maintaining fault tolerance~\cite{sunami2025transversalsurfacecodegamepowered}.

%%%%%
\textbf{Transversal H gate followed by a patch rotation.}
A logical H gate is implemented by applying physical H gates to all data qubits using Raman lasers (\cref{fig:transversal_gates}(b)).
For the rotated surface code, this swaps the code's boundary orientation.
To restore the original alignment required for subsequent transversal CX gates to act on the correct qubit pairs, the entire atom array must be physically rotated by $\pi/2$~\cite{Chen2026}.

%%%%%
\textbf{Teleported gates.}
The S and T gates can be implemented using dedicated resource states prepared by state distillation protocols~\cite{Bravyi2005,sunami2025transversalsurfacecodegamepowered}. 
These protocols use state injection~\cite{li2015magic}, transversal CX gates, and logical measurements~\cite{fowler2012surface,sunami2025transversalsurfacecodegamepowered}.
The prepared resource states are used for gate teleportation to implement the target logical gates, thereby allowing universal quantum computing~\cite{fowler2012surface,sunami2025transversalsurfacecodegamepowered}.

%%%%%%%%%%%%%%%%%%%%%%%%%%%%%%%%%%%%%
\subsection{Performance challenge and recent progress}

%%%
\subsubsection{Slow QEC with neutral atoms}
The widely adopted zoned architecture with global lasers results in ESM cycle times on the order of milliseconds, dominated by repeated atom shuttling.
By comparison, superconducting-qubit platforms complete an ESM round in $1~\mu\mathrm{s}$.
This gap has led to the widespread perception that neutral-atom FTQC is not competitive for practical applications.

%%%
\subsubsection{Recent progress towards fast QEC}
Two lines of recent progress are rapidly closing this gap.
On the hardware side, selective Rydberg gates and selective imaging (\cref{sec:atom_qc_overview}) enable shuttling-free ESM~\cite{sunami2025transversalsurfacecodegamepowered}, reducing the cycle time to hundreds of microseconds, limited primarily by ${\sim}100~\mu\mathrm{s}$ readout~\cite{falconi2025,sunami2025transversalsurfacecodegamepowered}.
On the theory side, advances in correlated decoding~\cite{cain2024correlated} and algorithmic fault tolerance~\cite{zhou2024algorithmic, cain2025fast} have shown that transversal-gate protocols can achieve threshold behavior with only $O(1)$ rounds of ESM per logical gate layer, rather than the conventional $d$ rounds, while maintaining polynomial-time decoding complexity~\cite{cain2025fast}.

\begin{figure}
	\centering
	\includegraphics[width=\figsize{}]{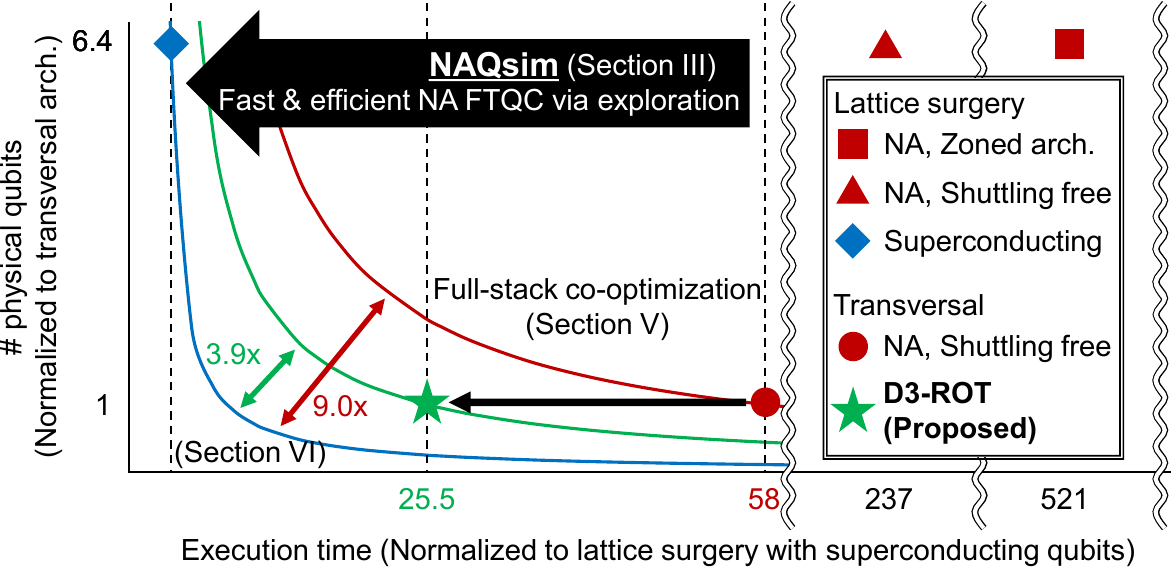}
	\caption{Our research goal and summary. Quantitative values are from our analysis in \cref{sec:evaluation}. The curves are contours of equal space-time cost.}
	\label{fig:perf_challenge}
\end{figure}

%%%%%%%%%%%%%%%%%%%%%%%%%%%%%%%%%%%%%
\subsection{Research challenges for realizing fast neutral-atom FTQC}

%%%
\subsubsection{Performance bottleneck in logical operations}
While ESM can be dramatically accelerated by selective operations, logical operations remain the critical bottleneck.
For example, transversal CX gates require AOD shuttling to overlap two logical-qubit patches before pairwise physical CX gates are applied. 
In addition, transversal H gates require rotation of the logical-qubit patches.
Even with abundant laser resources, these shuttling and rotation operations dominate the execution time, making it 58$\times$ longer than that of lattice surgery with superconducting qubits (``NA, Shuttling free'' in \cref{fig:perf_challenge}). 

%%%
\subsubsection{Complex interplay between FTQC software and hardware stacks}
Optimizing logical operations with NA systems is challenging due to the excessive flexibility available, as well as their complex operational constraints and error models.
Here, compiler-level decisions, such as gate scheduling, logical qubit placement, and routing, substantially affect the physical shuttling distances and available parallelism; 
conversely, hardware parameters such as the number of AOD channels, lattice spacing, and qubit plane geometry constrain compiler strategies.
These design choices involve fundamental trade-offs among space (physical qubit count and footprint), time (execution latency), and error (decoherence and gate infidelity).
A comprehensive co-design spanning the full stack is therefore essential to achieve high efficiency.

%%%
\subsubsection{Absence of a full-stack FTQC modeling tool for neutral-atom platforms}
To systematically explore this design space, architects need a full-stack FTQC modeling tool for neutral atoms that allows the evaluation of a given architecture's space, time, and error costs at the application level.
Such a tool must bridge from high-level fault-tolerant circuits down to physical laser operations and atom movements.
However, no such tool has been available: existing FTQC simulation frameworks target superconducting-qubit architectures~\cite{xqsim, qisim}, while Quration~\cite{suzuki2026quration} supports lattice-surgery-based FTQC beyond specific qubit technologies.
These tools do not capture the unique capabilities and constraints of neutral-atom platforms, such as reconfigurability, transversal gates, and shuttling latency.

%%%%%%%%%%%%%%%%%%%%%%%%%%%%%%%%%%%%%
\subsection{Research goal and overview}
In this paper, we address these challenges as outlined in \cref{fig:perf_challenge}.
First, we develop a full-stack simulation framework for neutral-atom FTQC (NAQsim) that models the transversal surface-code protocol (\cref{sec:3}). 
Next, we propose a near-rotation-free architecture (\archname{}) through full-stack co-optimization to resolve the previously overlooked performance bottleneck (\cref{sec:4}). 
Lastly, we provide extensive space-time overhead analysis in comparison with lattice surgery architectures to clearly illustrate our performance results (\cref{sec:evaluation}).

\begin{figure}
	\centering
	\includegraphics[width=\figsize{}]{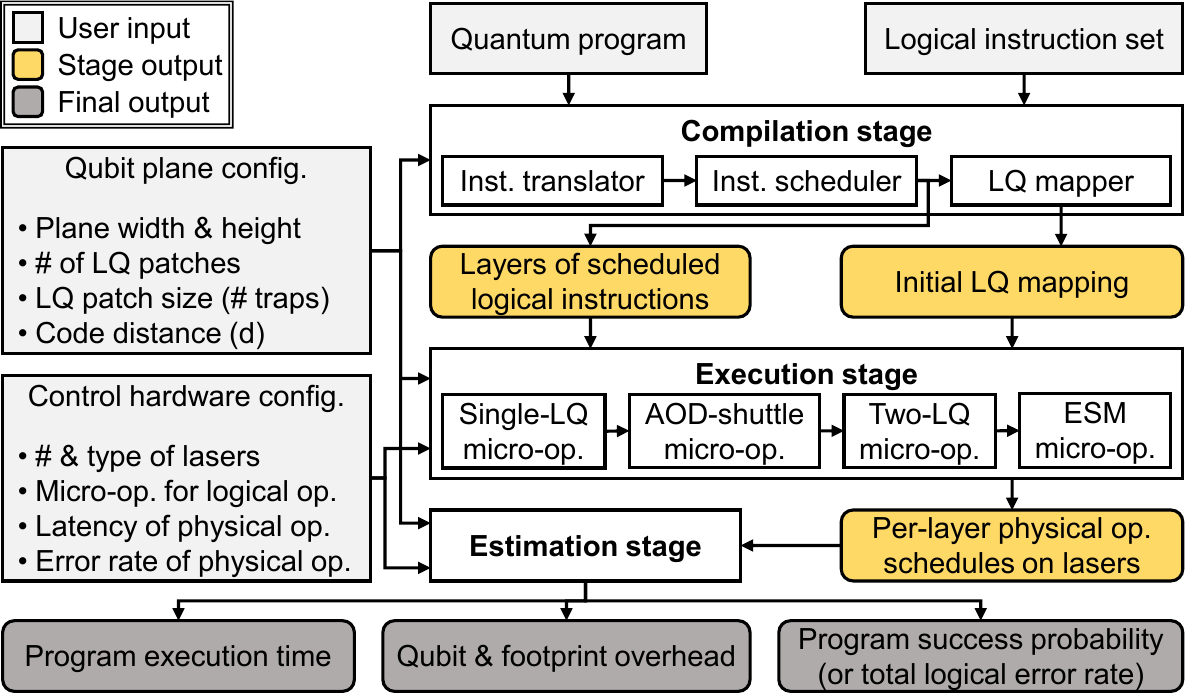}
	\caption{Overview of our simulation framework (NAQsim)}
	\label{fig:naqsim_overview}
\end{figure}

\section{NAQsim: Full-stack architecture simulation framework for FTQC with neutral atoms} \label{sec:3}

In this section, we describe NAQsim, our full-stack simulation framework for exploring various FTQC architectures of neutral-atom~(NA) quantum computers. 
We use the term ``full-stack'' to emphasize our tool's ability to evaluate the proposed ideas holistically across compiler, control hardware, and qubit plane stacks. 
\cref{fig:naqsim_overview} shows an overview of NAQsim, which proceeds through the compilation, execution, and estimation stages to derive the execution time, qubit and footprint overhead, and success rate of the target program.

%%%%%%%%%%%%%%%%%%%%%%%%%%%%%%%%%%%%%%%%%%%%%
\subsection{User inputs}

\begin{figure}
	\centering
	\includegraphics[width=\figsize{}]{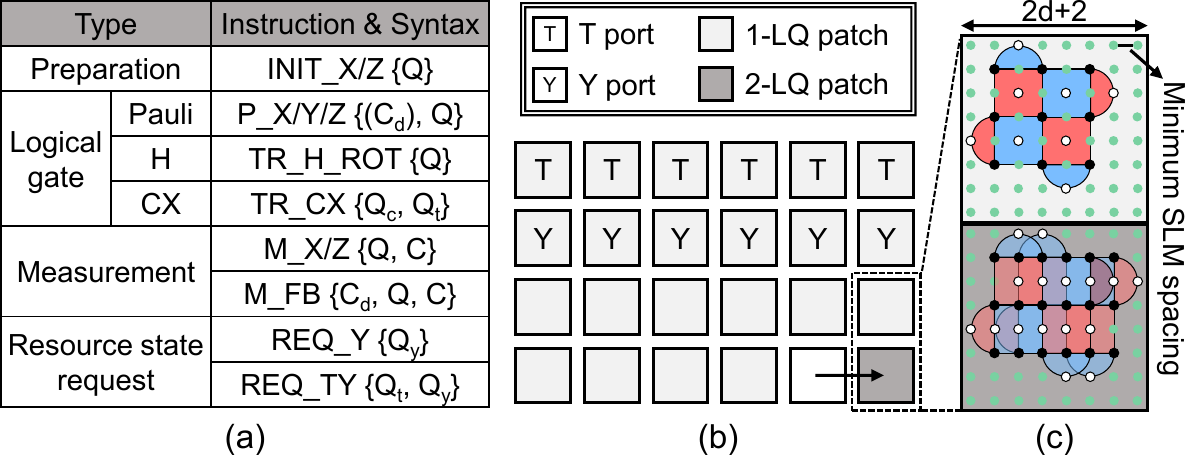}
	\caption{Baseline (a) logical instruction set and (b) qubit plane configuration with (c) the smallest logical-qubit (LQ) patch}
	\label{fig:isa_qbplane}
\end{figure}

We first introduce four major user inputs in detail.

%%%
\subsubsection{Quantum program} 
NAQsim can evaluate an arbitrary quantum program (or quantum circuit) built with the Clifford+T gate set, a universal gate set well suited for surface codes. 
We currently support quantum programs written in OpenQASM~\cite{openqasm2, openqasm3}. 

%%%
\subsubsection{Logical instruction set} \label{subsubsec:3.1.2}
\cref{fig:isa_qbplane}(a) shows our baseline logical instruction set and its syntax, where the symbols \texttt{Q} and \texttt{C} indicate the quantum and classical register operands, respectively. 
We introduce each logical instruction as follows. 

%%%%%
\textbf{Preparation.} 
The \texttt{INIT\_X} and \texttt{INIT\_Z} initialize the operand logical qubit (LQ) \texttt{Q} to $\ket{+}$ and $\ket{0}$, respectively.

%%%%%
\textbf{Logical gate.} 
We include three types of logical gate instructions to support the Clifford+T gate set with the transversal surface code. 
First, the \texttt{P\_X}, \texttt{P\_Y}, and \texttt{P\_Z} apply the corresponding Pauli gates to \texttt{Q}, optionally conditioned on the classical register $\texttt{C}_\texttt{d}$. 
Second, the \texttt{TR\_H\_ROT} applies the transversal H gate and a $\pi/2$ patch rotation to \texttt{Q}. 
Third, the \texttt{TR\_CX} implements the transversal CX gate for the control ($\texttt{Q}_\texttt{c}$) and target ($\texttt{Q}_\texttt{t}$) LQs. 

%%%%%
\textbf{Measurement.}
The \texttt{M\_X} and \texttt{M\_Z} measure \texttt{Q} in the X and Z bases, respectively, and then store the outcome in \texttt{C}. 
The \texttt{M\_FB} measures \texttt{Q} in the X or Z basis selected by a preceding measurement outcome $\texttt{C}_\texttt{d}$ and stores the result in \texttt{C}.

%%%%%
\textbf{Resource-state request.}
For S and T gate implementation, we use two resource states: $\ket{\mathrm{Y}}$ and $\ket{\mathrm{T}}$, prepared via distillation protocols. 
To separate the distillation protocol design from the compilation process, we introduce two resource-state request instructions following the philosophy in~\cite{lli_compiler}.
These instructions assume dedicated patches (or ports) for supplying resource states in the qubit plane, as shown in \cref{fig:isa_qbplane}(b). 
When a $\ket{\mathrm{Y}}$ state is ready, the \texttt{REQ\_Y} allocates a new quantum register $\texttt{Q}_\texttt{y}$.
Similarly, the \texttt{REQ\_TY} allocates $\texttt{Q}_\texttt{t}$ and $\texttt{Q}_\texttt{y}$ when both $\ket{\mathrm{T}}$ and $\ket{\mathrm{Y}}$ are ready. 
The specific ports used to supply resource states are selected during execution based on resource-state availability. 

%%%
\subsubsection{Qubit plane configuration} \label{subsubsec:3.1.3}
The qubit plane configuration defines the arrangement of LQ patches and each patch's structure.
\cref{fig:isa_qbplane}(b) shows our baseline qubit plane, which has the same number of patches as the input program's qubits, arranged in a nearly square shape. 
We additionally include two rows of patches for the Y and T ports above to handle the resource-state requests.

Each patch of our baseline employs the smallest-size configuration to minimize footprint overhead (\cref{fig:isa_qbplane}(c)). 
Specifically, the patch consists of $(2d+2)^2$ SLM traps to allow shuttling-free ESM over the patch with two overlapping LQs (or a 2-LQ patch). 
In addition, the SLM traps are separated by a minimum spacing (e.g., 5~$\mu\mathrm{m}$), such that at most one atom can shuttle between adjacent traps.

%%%
\subsubsection{Control hardware configuration}
The control hardware configuration includes the number and type of lasers, which significantly affect the implementation and latency of logical operations. 
The configuration also sets hardware parameters to define the latency and error rate of physical laser operations.

%%%%%%%%%%%%%%%%%%%%%%%%%%%%%%%%%%%%%%%%%%%%%
\subsection{Compilation stage}
The compilation stage generates two outputs: (1) layers of logical instructions and (2) an initial LQ mapping for the input quantum program and qubit plane configuration. 
To generate these outputs, the compilation stage consists of three components: an instruction translator, an instruction scheduler, and an LQ mapper. 
Note that our compilation is aware of the target FTQC protocol, but agnostic to the control hardware and physical qubit implementation. 

%%%
\subsubsection{Instruction translator} \label{subsubsec:3.2.1}

\begin{figure}
	\centering
	\includegraphics[width=\figsize{}]{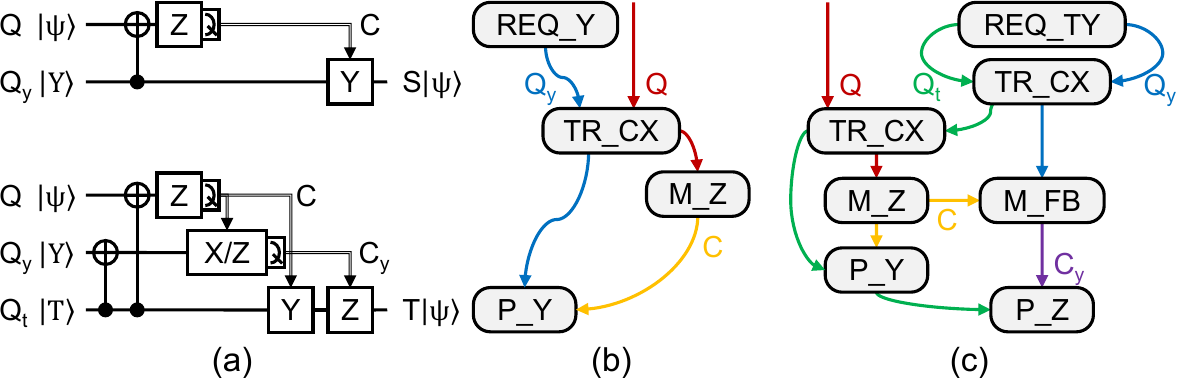}
	\caption{(a) Gate teleportation for logical S and T gates and our instruction translation for (b) S gate and (c) T gate}
	\label{fig:st_dag}
\end{figure}

First, the instruction translator converts the input program into a directed acyclic graph (DAG) representation using the given logical instruction set. 
In our DAG representation, the graph nodes and edges correspond to the instructions and operand dependencies, respectively.
\cref{fig:st_dag} shows an example of translation for S and T gates using the instructions in \cref{fig:isa_qbplane}(a). 
We adopt the S-gate and T-gate implementations from~\cite{sunami2025transversalsurfacecodegamepowered}, which utilize efficient gate teleportation using $\ket{\mathrm{Y}}$ and $\ket{\mathrm{T}}$ states.
As the program qubit's identifier changes with gate teleportation (e.g., from \texttt{Q} to $\texttt{Q}_\texttt{t}$ or $\texttt{Q}_\texttt{y}$), the instruction translator keeps track of the register identifiers and renames them as needed. 

%%%
\subsubsection{Instruction scheduler}
Second, using the translated DAG, the instruction scheduler iteratively generates layers of logical instructions. 
The term instruction layer indicates a list of scheduled instructions implicitly followed by ESM. 

%%%%%
\textbf{Scheduling \texttt{REQ\_Y} and \texttt{REQ\_TY}. }
The \texttt{REQ} instructions require careful consideration for their scheduling. 
Specifically, even when resource states are ready, we cannot schedule the corresponding \texttt{REQ} instructions when all the Y or T ports are in use. 
Therefore, we keep track of the available ports during scheduling (e.g., \texttt{TR\_CX} in gate teleportation frees a Y or T port). 
In addition, we schedule \texttt{REQ} instructions using As-Late-As-Possible (ALAP) ordering to prevent a deadlock.

%%%%%
\textbf{Scheduling other instructions. }
We skip the \texttt{P\_X}, \texttt{P\_Y}, and \texttt{P\_Z} instructions during scheduling as we can track these operations classically using a Pauli frame~\cite{knill2005quantum}. 
For other instructions, we basically adopt As-Soon-As-Possible (ASAP) scheduling. 

%%%
\subsubsection{Logical qubit mapper} \label{subsubsec:logical_qubit_mapper}
Third, the LQ mapper generates an initial LQ placement on the given qubit plane. 
For our baseline mapper that uses the instruction scheduler's output, we adopt the simulated annealing algorithm from the previous NISQ compiler~\cite{zac} and adapt it to our FTQC context. 
Specifically, we modify the moving cost function following \cref{eq:mov_cost}.
We also naturally support various simple initial mappings (e.g., row-major order mapping) as our baseline.

%%%%%%%%%%%%%%%%%%%%%%%%%%%%%%%%%%%%%%%%%%%%%
\subsection{Execution stage}

\begin{figure}
	\centering
	\includegraphics[width=\figsize{}]{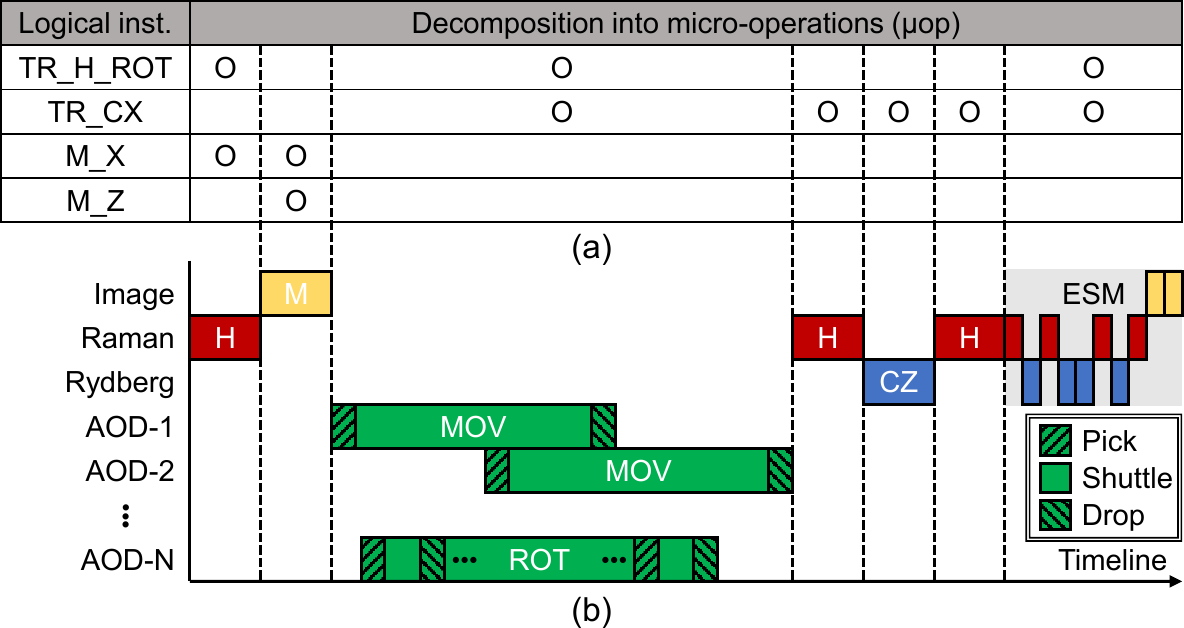}
	\caption{(a) Decomposition of instructions into laser-wise \textmu{}ops and (b) example execution trace of one instruction layer (\textmu{}op\_ prefix omitted)}
	\label{fig:uop_timeline}
\end{figure}

Starting from the initial mapping, the execution stage iteratively generates an execution trace of every scheduled instruction layer.
The execution trace records the detailed timeline of physical laser operations and their operand LQ(s). 
To obtain the trace, the execution stage (1) decomposes instructions into laser-wise micro-operations (\textmu{}ops), (2) simulates their execution with accurate latency modeling, and (3) updates the LQs' status and location.

%%%
\subsubsection{Laser-wise micro-operations}
In general, FTQC instructions are specified per LQ and logical operation, agnostic to control hardware characteristics. 
However, in NA quantum computers, a single laser can simultaneously apply physical operations to all target physical qubits. 
In addition, each logical operation needs multiple physical operations, each of which may require a different type of laser. 

To address this issue, we introduce the concept of a micro-operation~(\textmu{}op), similar to how an x86 machine translates a complex instruction into multiple simple, hardware-aware operations. 
In our case, a \textmu{}op represents a set of physical quantum operations supported by a single laser type, each of which can be applied to multiple LQs simultaneously. 
\cref{fig:uop_timeline} shows how our instructions are decomposed into \textmu{}ops and executed. 

%%%%%
\textbf{\textmu{}op\_H with Raman laser.}
A single Raman laser applies physical H gates to all data qubits of multiple LQs.

%%%%%
\textbf{\textmu{}op\_M with imaging laser.}
A single imaging laser measures all data (or ancilla) qubits of multiple LQs. 

%%%%%
\textbf{\textmu{}op\_CZ with Rydberg laser.}
A single Rydberg laser applies CZ gates to all target physical qubit pairs selectively. 

%%%%%
\textbf{\textmu{}op\_MOV with AOD laser.}
The move \textmu{}op with an AOD laser is a key feature of NA quantum computing. 
In detail, the move consists of three physical operations: picking up atoms from SLM traps, shuttling them from the source to the destination patch, and dropping them into destination SLM traps.
This \textmu{}op uses a single AOD laser and can move multiple LQs unless there is an AOD conflict. 

%%%%%
\textbf{\textmu{}op\_ROT with AOD laser.}
The patch rotation \textmu{}op is the AOD operation necessary to rotate LQ(s). 
Most importantly, there are various implementations of the \textmu{}op\_ROT, depending on AOD type and qubit plane configuration. 
For example, the patch rotation can consist of multiple atom movements (e.g., \cref{fig:uop_timeline}(b)) or not.
We will explore the trade-offs of its implementation in detail (\cref{subsec:4.3}).

%%%%%
\textbf{Error syndrome measurement.}
At the end of each instruction layer, we perform one round of ESM for all the live LQs (i.e., not yet destructively measured). 
\cref{fig:uop_timeline} shows how we add the ESM round with a sequence of \textmu{}op\_H, \textmu{}op\_CZ, and \textmu{}op\_M.

%%%
\subsubsection{Simulation pipeline of AOD \textmu{}ops} \label{subsubsec:3.3.2}

\begin{figure}
	\centering
	\includegraphics[width=\figsize{}]{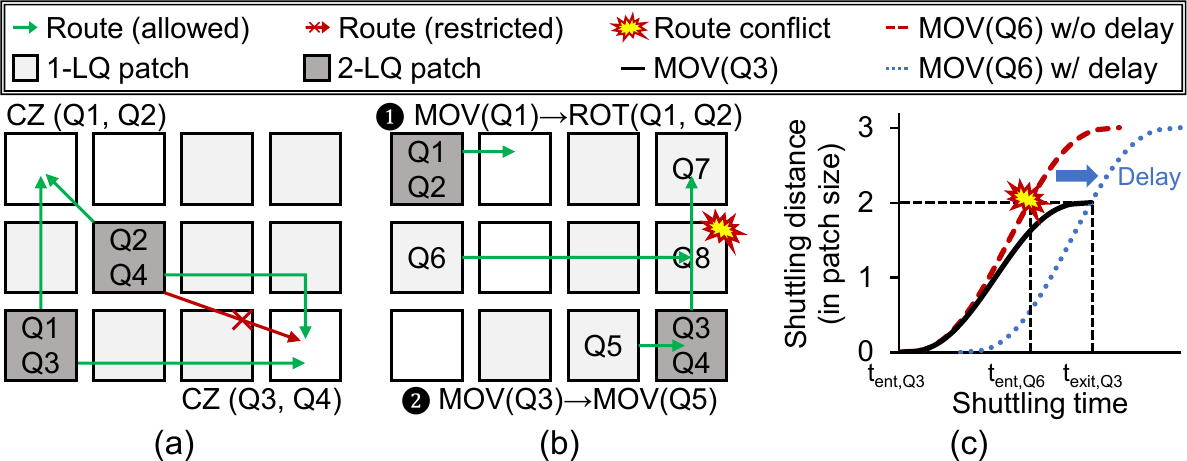}
	\caption{Examples of (a) restricted moving routes, (b) LQ and patch dependencies, and (c) route conflicts during AOD \textmu{}op execution}
	\label{fig:aod_pipeline}
\end{figure}

Notably, the most complex part of our execution stage is the detailed simulation of AOD \textmu{}ops. 
We describe our modeling methodology, inspired by the conventional processor pipeline workflow.

%%%%%
\textbf{Decode.}
First, we decode logical instructions and generate AOD \textmu{}ops. 
For example, it is straightforward to generate \textmu{}op\_ROT for the same operand LQ as the \texttt{TR\_H\_ROT}. 
Generating \textmu{}op\_MOV is much more complex. 
The major source of \textmu{}op\_MOV is the \texttt{TR\_CX} instruction, whose two operands should be in the same patch before CX is applied.
As the simplest way to satisfy this requirement, we can move one LQ to the other's patch. 
However, this is often impossible when both operand LQs are already in different 2-LQ patches. 

For systematic move generation, we adopt the CX placement approach proposed in~\cite{zac}. 
While the original idea focuses on NISQ zoned architectures, we observe that our LQ patch is conceptually similar to the physical Rydberg site (i.e., both require two CX operands to be brought to the same location).
In summary, we employ a minimum-weight full matching on a bipartite graph, where the two vertex sets (left and right) are the \texttt{TR\_CX} instructions and the candidate LQ patches (i.e., all except 2-LQ patches). 
An edge between the left and right vertices has a weight equal to the sum of the costs of the required moves.

As a key difference in the FTQC, we should account for the restricted moving routes. 
For example, we can collectively move all the physical qubits of an LQ in the horizontal, vertical, and diagonal directions (e.g., green arrows in \cref{fig:aod_pipeline}(a)). 
Moving in other directions incurs collisions between AOD-trap atoms and SLM traps (with atoms) due to the regular SLM spacing (e.g., red arrow in \cref{fig:aod_pipeline}(a)).
Therefore, in the FTQC context, we should divide these moves into two steps in the allowed directions (e.g., MOV(Q4) in \cref{fig:aod_pipeline}(a)). 
We redefine the moving cost function for our CX placement to account for the restricted route problem:
\begin{align}
\label{eq:mov_cost}
&\operatorname{MovingCost}(\mathrm{src}, \mathrm{dst}) = 
\begin{cases}
\sqrt[3]{\sqrt{x_{\mathrm{diff}}^2 + y_{\mathrm{diff}}^2}} & \text{for one-step}\\
\sqrt[3]{x_{\mathrm{diff}}} + \sqrt[3]{y_{\mathrm{diff}}} & \text{for two-step}
\end{cases}, 
\end{align}
where $x_{\mathrm{diff}}$ equals $|\mathrm{src}_x-\mathrm{dst}_x|$ and $y_{\mathrm{diff}}$ equals $|\mathrm{src}_y-\mathrm{dst}_y|$.
The cube root of the distance originates from our shuttle latency estimation based on STA trajectories~\cite{hwang2024fast}.

%%%%%
\textbf{Dependency check.}
Second, we analyze two types of dependencies among AOD \textmu{}ops: LQ and patch dependencies. 
LQ dependency exists between two \textmu{}ops on the same patch. 
For example, to rotate both LQs at the same patch, we need to move one LQ to another patch and then rotate them (e.g., \cref{fig:aod_pipeline}(b)\ding{182}). 
Patch dependency exists between two \textmu{}ops if one leaves the patch, and the other enters it (or takes it up) (e.g., \cref{fig:aod_pipeline}(b)\ding{183}).  
Based on our dependency analysis, we build a DAG of the AOD \textmu{}ops.

%%%%%
\textbf{Issue.}
Third, we iteratively issue an AOD \textmu{}op with wake-up and select stages.
We first wake up all the ready \textmu{}ops that are free of dependencies and structural hazards (i.e., the required number of AOD lasers is available).
Then, we select one \textmu{}op for execution based on the selection priority policy. 
In this work, we basically prioritize the first-finishing \textmu{}op based on latency estimation.

%%%%%
\textbf{Execute.}
Finally, we simulate the execution of the selected \textmu{}op and estimate the detailed timeline of pick, shuttle, and drop operations.
For accurate latency estimation in NA FTQC, we introduce a novel concept we term ``route conflict''.
\cref{fig:aod_pipeline}(b) shows a route conflict example for MOV(Q3) and MOV(Q6). 
The two moves can start simultaneously as they are free of any dependencies; however, we should prevent both Q3 and Q6 from simultaneously passing through the same patch, e.g., the Q8 patch. 
Otherwise, their physical atoms can collide, significantly affecting LERs.  
Naively waiting until one move finishes can substantially overestimate the execution time.

To address this challenge, we analyze detailed time intervals for the patches touched during \textmu{}op execution.
As \cref{fig:aod_pipeline}(c) shows, when we issue MOV(Q3) first, the atoms of Q3 begin entering the Q8 patch at $t_{\mathrm{ent}, \texttt{Q3}}$ and fully exit the patch at $t_{\mathrm{exit}, \texttt{Q3}}$.
On the other hand, when MOV(Q6) also starts at the same time, Q6 enters the Q8 patch at $t_{\mathrm{ent}, \texttt{Q6}}$, thereby causing a route conflict. 
With this analysis, we can accurately determine the minimum delay required to avoid a route conflict (i.e., $t_{\mathrm{exit}, \texttt{Q3}} - t_{\mathrm{ent}, \texttt{Q6}}$ in \cref{fig:aod_pipeline}(c)). 
We iterate this analysis for every \textmu{}op, derive the execution timeline with the obtained delay, and record the time intervals for the touched patches.

After completing the issued \textmu{}op's simulation,  we update the execution trace and remove the \textmu{}op from the DAG. 
We iterate on the issue and execute steps until the DAG drains, indicating the end of simulation for one instruction layer.

%%%%%%%%%%%%%%%%%%%%%%%%%%%%%%%%%%%%%%%%%%%%%
\subsection{Estimation stage}
Based on the execution traces and input configurations, the estimation stage derives the total execution time, qubit and footprint overhead, and program success rate. 

%%%
\subsubsection{Program execution time}
We can naturally derive the total execution time from our execution trace, which captures the timeline of all physical operations. 
More importantly, we can analyze performance bottlenecks with a detailed execution-time breakdown, an essential capability for architecture exploration and optimization.

%%%
\subsubsection{Qubit \& footprint overhead}
With the qubit-plane configuration, we calculate the number of physical qubits as $\text{(\# LQ patches)} \times{} (2d^2-1)$. 
In addition, we derive the footprint overhead (i.e., the area of the qubit plane) to quantify the overhead of ideas that require larger patch sizes for the same number of physical qubits. 

%%%
\subsubsection{Program success rate}\label{subsubsec:success_probability}

\begin{figure}
	\centering
	\includegraphics[width=\figsize{}]{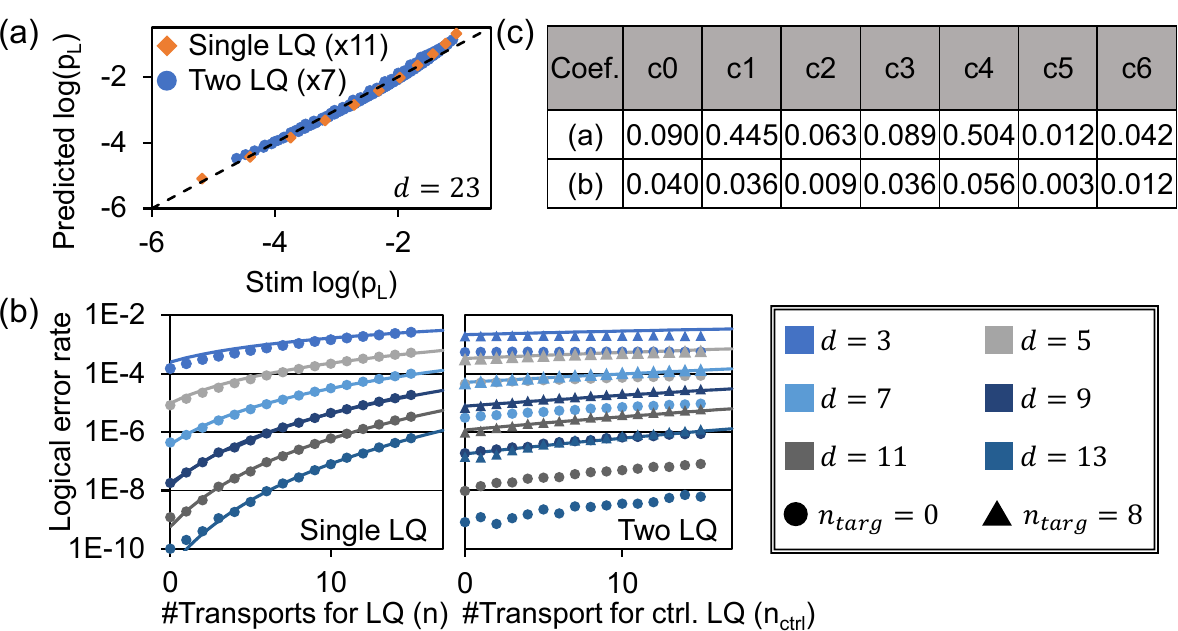}
	\caption{(a) LER model validation result and (b) LER modeling plots. (c) Coefficient values for (a) and (b).}
	\label{fig:logical_error_model}
\end{figure}

Based on the execution trace, NAQsim estimates the program success rate by accumulating LERs of all live LQs layer by layer.
A unique feature of our error estimation is the incorporation of AOD--SLM transfer noise; in NA platforms, moving an LQ requires picking (dropping) atoms from (to) the SLM by using AODs~\cite{Cicali2026prappl}.
We model this transfer penalty by applying depolarizing noise to all physical qubits in the LQ for each pick/drop.

To efficiently evaluate the LER per layer, we adopt the fast correlated decoding framework~\cite{cain2025fast}, which allows us to maintain fault tolerance with $O(1)$ ESM rounds per transversal gate.
We classify the LQ operations in a layer into two categories: 1-LQ operations and 2-LQ transversal CX operations.
We approximate the LER of a 1-LQ operation by using the LER of a single-round memory operation, and evaluate the LER of a 2-LQ operation based on the correlated decoding approach~\cite{cain2025fast}. 

\begin{align} 
    & P_{\mathrm{L},\ \text{1Q}}(d,\ n) = c_0 (c_1+c_2n)^{d / 2} \label{eq:LER_model_1q} \\ 
    & P_{\mathrm{L},\ \text{2Q}}(d,\ n_\text{ctrl},\ n_\text{targ}) = c_3 (c_4+c_5n_\text{ctrl} + c_6n_\text{targ})^{d / 2} \label{eq:LER_model_2q} 
\end{align}

Using Stim~\cite{gidney2021stim} and PyMatching~\cite{higgott2022pymatching}, we model the LER as a function of the code distance $d$ and the number $n$ of AOD transfers per layer.
For transversal CXs, we track the pick/drop counts for both the control ($n_{\text{ctrl}}$) and target ($n_{\text{targ}}$) LQs. 
We set other error parameters following \cref{tab:eval_setup}.
By fitting the simulation results across various $d$ and $n$ values, we derive analytical models for 1-LQ operations in \cref{eq:LER_model_1q} and 2-LQ operations in \cref{eq:LER_model_2q}, where $c_0$ to $c_6$ are constants.

We validate our LER models against Stim simulations at $d=23$ by scaling the error parameters in \cref{tab:eval_setup} ($11\times$ for 1-LQ and $7\times$ for 2-LQ) to ensure the $d=23$ LERs can be measured directly with sufficient samples. 
We fit the models using $d = 3,\ldots,13$ data and compare the predictions against the $d=23$ Stim simulations while sweeping $n$,  $n_\text{ctrl}$, and $n_\text{targ}$.
\cref{fig:logical_error_model}(a) shows that the fitted models match the Stim simulation results well, with RMSEs in log10 LER of 0.162 and 0.097 for 1-LQ and 2-LQ operations, respectively.

After the validation, we apply the same fitting procedure to the target error setting in \cref{tab:eval_setup}.
\cref{fig:logical_error_model}(b) shows the simulation results as a scatter plot and the model predictions as lines. 
\cref{fig:logical_error_model}(c) reports the fitted coefficients for the settings in \cref{fig:logical_error_model}(a) and (b).

\section{Evaluation methodology}

\begin{table}[t]
\caption{Evaluation setup}
\centering
\tabcolsep=3.0pt
%\resizebox{\columnwidth}{!}{%
\scriptsize
\begin{tabular}{|ccccc|}
\hline
\rowcolor[HTML]{AFABAB} 
\multicolumn{1}{|c|}{\cellcolor[HTML]{AFABAB}\begin{tabular}[c]{@{}c@{}}Quantum program\end{tabular}} &
  \multicolumn{2}{c|}{\cellcolor[HTML]{AFABAB}\begin{tabular}[c]{@{}c@{}}LQ plane size\end{tabular}} &
  \multicolumn{2}{c|}{\cellcolor[HTML]{AFABAB}\begin{tabular}[c]{@{}c@{}}Min. LQ patch side\\($L_{\mathrm{min}}$)\end{tabular}} \\ \hline
\multicolumn{1}{|c|}{} &
  \multicolumn{1}{c|}{Width \& Height} &
  \multicolumn{1}{c|}{T \& Y ports} &
  \multicolumn{1}{c|}{Trap} &
  Spacing \\ \cline{2-5} 
\multicolumn{1}{|c|}{\multirow{-2}{*}{\cref{tab:benchmark_programs}}} &
  \multicolumn{1}{c|}{\begin{tabular}[c]{@{}c@{}}\# LQs in a\\ nearly square shape\end{tabular}} &
  \multicolumn{1}{c|}{1 row each} &
  \multicolumn{1}{c|}{2$d$+2} &
  5~$\mu\mathrm{m}$ \\ \hline
\rowcolor[HTML]{AFABAB} 
\multicolumn{5}{|c|}{\cellcolor[HTML]{AFABAB}Control hardware configuration} \\ \hline
\multicolumn{1}{|c|}{Laser} &
  \multicolumn{1}{c|}{Operation} &
  \multicolumn{1}{c|}{Latency ($\mu\mathrm{s}$)} &
  \multicolumn{1}{c|}{Error rate} &
  \# lasers \\ \hline
\multicolumn{1}{|c|}{} &
  \multicolumn{1}{c|}{DRO} &
  \multicolumn{1}{c|}{60~\cite{ma2023high-fidelity}} &
  \multicolumn{1}{c|}{1e-3} &
   \\ \cline{2-4}
\multicolumn{1}{|c|}{\multirow{-2}{*}{Imaging}} &
  \multicolumn{1}{c|}{NDRO} &
  \multicolumn{1}{c|}{100~\cite{falconi2025}} &
  \multicolumn{1}{c|}{1e-3} &
  \multirow{-2}{*}{2} \\ \hline
\multicolumn{1}{|c|}{Pump} &
  \multicolumn{1}{c|}{Reset} &
  \multicolumn{1}{c|}{100~\cite{Norcia2023}} &
  \multicolumn{1}{c|}{1e-3} &
  1 \\ \hline
\multicolumn{1}{|c|}{Raman} &
  \multicolumn{1}{c|}{1Q (H)} &
  \multicolumn{1}{c|}{1~\cite{evered2023high-fidelity,jenkins2022ytterbium}} &
  \multicolumn{1}{c|}{1e-4} &
  1 \\ \hline
\multicolumn{1}{|c|}{} &
  \multicolumn{1}{c|}{CZ} &
  \multicolumn{1}{c|}{1~\cite{Radnaev2025}} &
  \multicolumn{1}{c|}{1e-3} &
   \\ \cline{2-4}
\multicolumn{1}{|c|}{\multirow{-2}{*}{Rydberg}} &
  \multicolumn{1}{c|}{\begin{tabular}[c]{@{}c@{}}Pattern change\end{tabular}} &
  \multicolumn{1}{c|}{1~\cite{Radnaev2025, Graham2023multiscale}} &
  \multicolumn{1}{c|}{-} &
  \multirow{-2}{*}{1$\sim$2} \\ \hline
\multicolumn{1}{|c|}{} &
  \multicolumn{1}{c|}{\begin{tabular}[c]{@{}c@{}}Pick / Drop\end{tabular}} &
  \multicolumn{1}{c|}{40~\cite{Cicali2026prappl}} &
  \multicolumn{1}{c|}{1e-3} &
   \\ \cline{2-4}
\multicolumn{1}{|c|}{\multirow{-2}{*}{AOD}} &
  \multicolumn{1}{c|}{Shuttle} &
  \multicolumn{1}{c|}{STA~\cite{hwang2024fast}} &
  \multicolumn{1}{c|}{T2 = 10~$\mathrm{s}$} &
  \multirow{-2}{*}{Sufficient} \\ \hline
\multicolumn{1}{|c|}{} &
  \multicolumn{1}{c|}{\begin{tabular}[c]{@{}c@{}}Pattern change\end{tabular}} &
  \multicolumn{1}{c|}{1000~\cite{Lin2025aienabled}} &
  \multicolumn{1}{c|}{-} &
   \\ \cline{2-4}
\multicolumn{1}{|c|}{\multirow{-2}{*}{SLM}} &
  \multicolumn{1}{c|}{\begin{tabular}[c]{@{}c@{}}Turn on/off\end{tabular}} &
  \multicolumn{1}{c|}{1~\cite{bluvstein2024logical, Graham2023multiscale}} &
  \multicolumn{1}{c|}{-} &
  \multirow{-2}{*}{1$\sim$2} \\ \hline
\end{tabular}%
%}
\label{tab:eval_setup}
\end{table}

%%%%%%%%%%%%%%%%%%%%%%%%%%%%%%%%%%%%%%%%%%%%%
\subsection{FTQC benchmark workloads}
We use eight practical FTQC workloads spanning four representative algorithm classes with various logical-qubit scales and program sizes. 
For the QPE workloads, we utilize Quration~\cite{suzuki2026quration} to generate Clifford+T circuits for the target Hamiltonians.
For other workloads, we utilize OpenQASM circuits from FTCircuitBench~\cite{ftcircuitbench} and transpile them with Qiskit, synthesizing arbitrary-angle rotations into Clifford+T with a program-level synthesis-error bound of $10^{-3}$.
\cref{tab:benchmark_programs} summarizes the circuit statistics for our target workloads.

%%%
\subsubsection{Quantum phase estimation (QPE)}
QPE estimates the ground-state energy of a target Hamiltonian and is a representative early-FTQC application~\cite{babbush2018encoding,reiher2017elucidating,yoshioka2022hunting,kivlichan2020improved,lee2021even}. 
We evaluate the \texttt{SELECT} subroutine of qubitization-based QPE~\cite{low2019hamiltonian}, which dominates the execution time for typical Hamiltonians relevant to early applications~\cite{yoshioka2022hunting}. 
Following prior QPE resource-estimation studies~\cite{yoshioka2022hunting,kivlichan2020improved,lee2021even,babbush2018encoding}, we use four Hamiltonians: 2D Heisenberg~(Hei) model, 2D Fermi--Hubbard~(FH) model, Jellium, and $\mathrm{H}_4$ molecular system~(H4), covering local spin, local fermionic, long-range electronic structure, and small molecular systems, respectively.
We evaluate a single \texttt{SELECT} operation and scale the total cost by the repetition count determined from the Hamiltonian norm and the target energy accuracy of $0.01$.

\begin{table}[tb]
\tabcolsep=2.0pt
\scriptsize
\caption{FTQC benchmark workloads}
\label{tab:benchmark_programs}
%\resizebox{\columnwidth}{!}{%
\begin{tabular}{|c|c|c|c|c|c|c|c|c|} \hline
\rowcolor[HTML]{AFABAB} 
Class              & \multicolumn{4}{c|}{Quantum phase estimation} & Arith. & QFT    & \multicolumn{2}{c|}{
\begin{tabular}[c]{@{}c@{}}Hamiltonian\\simulation\end{tabular}} \\ \hline
\begin{tabular}[c]{@{}c@{}}Workload\end{tabular}& 
\begin{tabular}[c]{@{}c@{}}Hei\end{tabular} &
\begin{tabular}[c]{@{}c@{}}FH\end{tabular}& 
\begin{tabular}[c]{@{}c@{}}Jellium\end{tabular}& 
\begin{tabular}[c]{@{}c@{}}H4\end{tabular}& 
\begin{tabular}[c]{@{}c@{}}adder\end{tabular}& 
\begin{tabular}[c]{@{}c@{}}qft\end{tabular}& 
\begin{tabular}[c]{@{}c@{}}ising1d\end{tabular}& 
\begin{tabular}[c]{@{}c@{}}ising2d\end{tabular} \\ \hline
\#LQs              &167        &95       &41           &27         &64          &29           &64              &64           \\
\#Repetitions      &48067      &56549    &38075        &975        &-           &-            &-               &-            \\
\#CX gates             &32414      &43125    &71134        &8778       &455         &722          &640             &1280         \\
\#H gates         &18370      &12195    &22424        &3404       &12376       &48737        &39680           &58880        \\
\#S gates          &11886      &7521     &14700        &2246       &5908        &25552        &20800           &31680        \\
\#T gates          &12968      &9324     &17352        &2616       &12264       &48236        &39040           &58240        \\ \hline
Trans. $d$           &23         &23       &23           &17         &13          &15           &15              &17           \\
Trans. \#qubits    &177576     &104643   &44394        &17310      &21568       &13470        &28736           &36928        \\
LS $d$             &  27       & 27      & 27          & 21        & 17         & 19          & 19             & 21          \\
LS \#qubits        &1014072    &603198   &262260       &114530     &156944      &93730        &196112          &239632       \\ \hline
\end{tabular}
%}
\end{table}

%%%
\subsubsection{Hamiltonian simulation}
Hamiltonian simulation is a core primitive for studying quantum dynamics and implements time evolution under a target Hamiltonian.
We evaluate the 5-Trotter-step 1D and 2D Ising-model circuits from~\cite{ftcircuitbench}.
Unlike QPE, these circuits directly implement time-evolution kernels; the 1D/2D pair captures different interaction graphs and locality patterns.

%%%
\subsubsection{Quantum adder}
Quantum addition is a basic reversible-arithmetic primitive used in modular arithmetic, period finding, and arithmetic oracles.
We include the adder benchmark from FTCircuitBench~\cite{ftcircuitbench} to cover arithmetic-dominated workloads.

%%%
\subsubsection{Quantum Fourier Transform (QFT)}
QFT maps computational-basis states to the Fourier basis and is a standard subroutine in phase estimation, period finding, and arithmetic routines.
We include the QFT benchmark from~\cite{ftcircuitbench} to cover circuits with regular long-range controlled-rotation patterns.
The QFT complements the local Hamiltonian-simulation and carry-propagation workloads by stressing nonlocal logical interactions and rotation synthesis.

%%%%%%%%%%%%%%%%%%%%%%%%%%%%%%%%%%%%%%%%%%%%%
\subsection{Qubit plane and control hardware setup}

%%%
\subsubsection{Qubit plane}
We mainly follow the baseline configuration described in \cref{subsubsec:3.1.3}.
We start our exploration with a qubit plane composed of the smallest LQ patches with the minimum side length $L_{\mathrm{min}}$ (\cref{tab:eval_setup}). 
We set the code distance $d$ for each workload to achieve a program success rate of 99.9\% (summarized in \cref{tab:benchmark_programs}). 

%%%
\subsubsection{Control hardware}
We adopt the state-of-the-art selective laser operations and their latency values for shuttling-free ESM~\cite{ma2023high-fidelity, falconi2025, Norcia2023, evered2023high-fidelity, jenkins2022ytterbium, Radnaev2025, Graham2023multiscale, Cicali2026prappl, hwang2024fast, Lin2025aienabled, bluvstein2024logical}. 
We assume a physical error rate of 0.1\% for CZ gates, measurement, reset, and AOD--SLM transfers, and 0.01\% for single-qubit gates (\cref{tab:eval_setup}).
Although NAQsim can also model scenarios with fewer AODs, we assume that a sufficiently large number of AODs are available. 
We intentionally focus on this scenario to highlight critical performance bottlenecks that cannot be addressed by simply adding more lasers.
We provide a sensitivity analysis under limited AOD resources in \cref{subsec:sensitivity}. 

\section{Near-rotation-free FTQC architecture design with full-stack co-optimization} \label{sec:4}

This section demonstrates a use case of NAQsim to explore FTQC architectures. 
Driven by NAQsim, we propose a near-rotation-free architecture with compiler, control hardware, and qubit plane co-optimization for fast, space-efficient FTQC.

%%%%%%%%%%%%%%%%%%%%%%%%%%%%%%%%%%%%%%%%%%%%%%%%
\subsection{Rotation-induced performance bottleneck}

\begin{figure}
	\centering
	\includegraphics[width=\figsize{}]{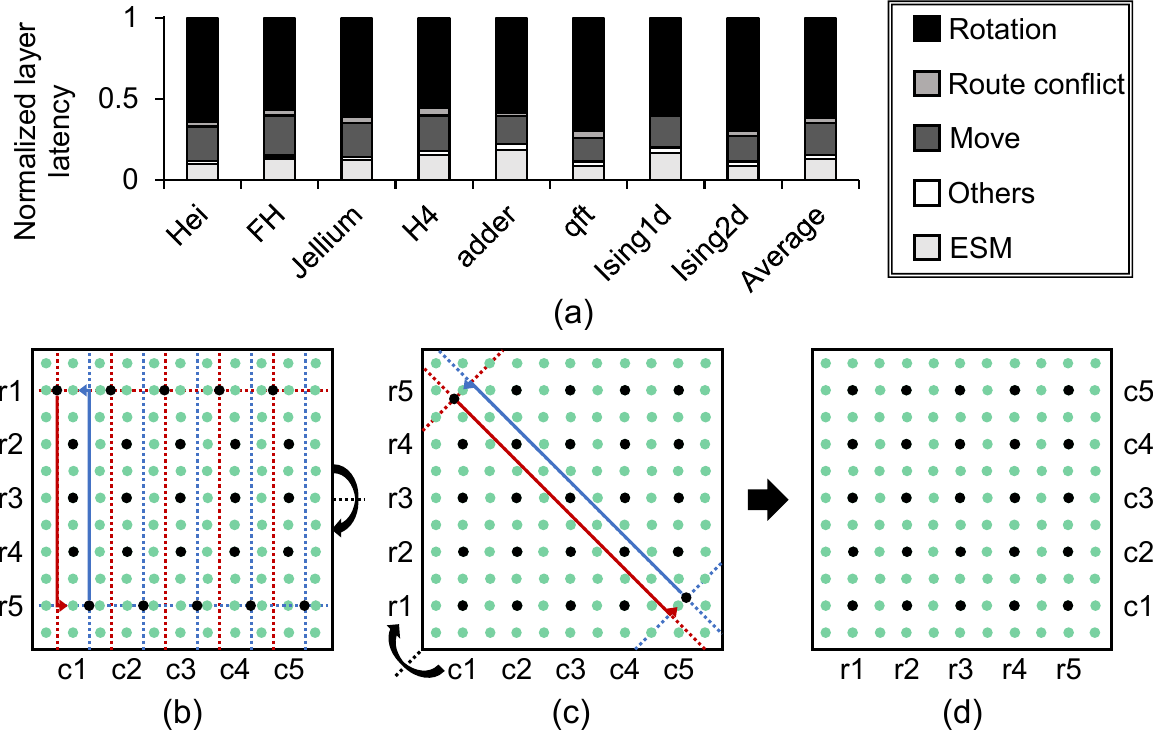}
	\caption{(a) Performance bottleneck analysis; logical-qubit patch rotation through space-efficient reflection: row-by-row (b) horizontal and (c) diagonal swaps resulting in (d) the rotated patch.} 
	\label{fig:rot_bottleneck}
\end{figure}

Driven by NAQsim, we observe that the logical-qubit patch rotation is the critical bottleneck in the baseline architecture.
\cref{fig:rot_bottleneck}(a) breaks down the AOD execution time into the move latency (Move), additional delay due to route conflicts (Route conflict), and exposed patch rotation latency (Rotation). 
Notably, the patch rotation dominates the execution time for all FTQC workloads, accounting for 61.9\% on average.

\cref{fig:rot_bottleneck}(b)--(d) elaborate on the bottleneck analysis by illustrating the implementation based on space-efficient reflection. 
In principle, we perform LQ rotation in two steps: patch reflections along the horizontal (\cref{fig:rot_bottleneck}(b)) and diagonal axes (\cref{fig:rot_bottleneck}(c)).
However, patch reflection within the smallest patch requires sequential swaps of atom rows, which significantly increases execution time.

%%%%%%%%%%%%%%%%%%%%%%%%%%%%%%%%%%%%%%%%%%%%%%%%
\subsection{Compiler optimization\label{subsec:compilier_opitimization}}

We first explore optimization opportunities at the compilation stage. 
Previous FTQC studies assume that the patch rotation should always follow the logical H operation to recover the boundary shape.
However, the real constraint that must be satisfied in the transversal architecture is that two operands of a transversal CX should have the same boundary shape.
Thus, we aim to reduce rotation overhead by focusing on the gap between current practice and the true constraint. 

%%%
\subsubsection{Opportunities exposed by separate ROT instruction}

\begin{figure}
	\centering
	\includegraphics[width=\figsize{}]{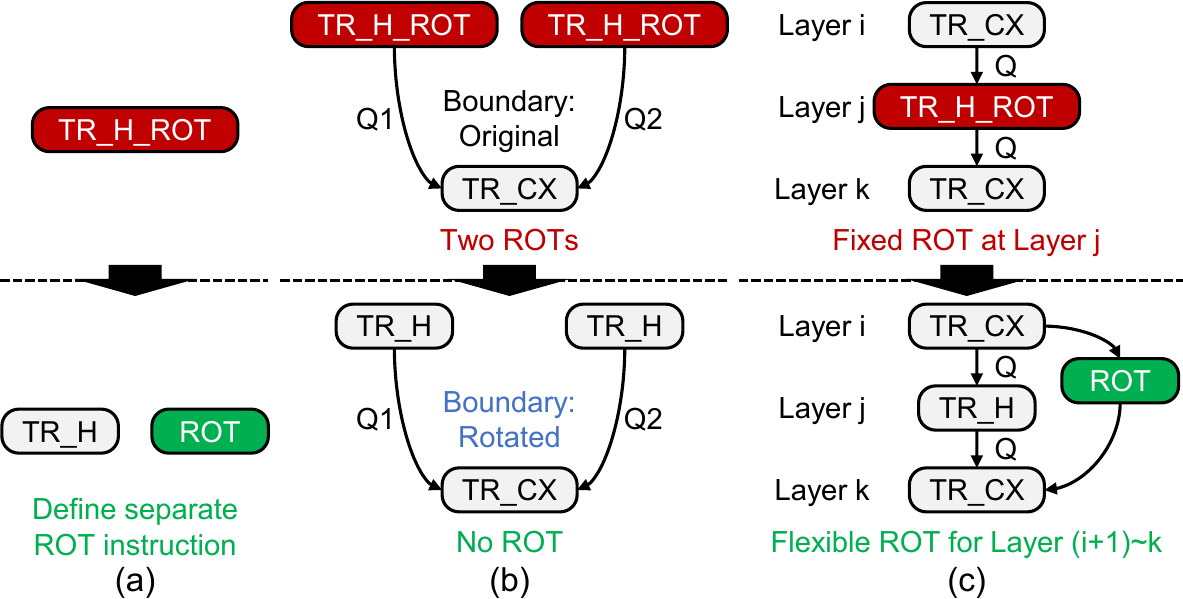}
	\caption{Compiler optimization opportunities: (a) decoupling the \texttt{ROT} instruction from \texttt{TR\_H\_ROT}, (b) skipping unnecessary rotations, and (c) enabling flexible scheduling of \texttt{ROT}.}
	\label{fig:compiler_idea}
\end{figure}

We first separately define the logical instructions for transversal H (\texttt{TR\_H}) and patch rotation (\texttt{ROT}) (\cref{fig:compiler_idea}(a)).
With this separate \texttt{ROT} instruction, we expose two novel compiler optimization opportunities.

\textbf{Opportunity \#1: unnecessary patch rotations.} 
Separating \texttt{TR\_H\_ROT} allows us to identify opportunities to skip patch rotations by considering the real constraint on the boundary shape. 
For example, if both operands of a transversal CX have rotated boundaries, we do not have to rotate them (\cref{fig:compiler_idea}(b)). 

\textbf{Opportunity \#2: flexible rotation scheduling.} 
While \texttt{TR\_H\_ROT} enforces that the transversal H and rotation be in the same layer, separating the instruction provides much more flexibility for the timing of rotation. 
For example, \cref{fig:compiler_idea}(c) shows that we can now schedule the \texttt{ROT} instruction in any layer between the \texttt{TR\_CX} instructions preceding and succeeding the \texttt{TR\_H}.

%%%
\subsubsection{Optimization \#1. Rotation skip in instruction translator}
To exploit Opportunity \#1, we introduce a rotation skip algorithm into the instruction translator. 
First, we construct an instruction DAG without \texttt{ROT} instructions and analyze the expected boundary shapes of the operands of every \texttt{TR\_CX}. 
Next, we extract a subgraph of \texttt{TR\_CX} instructions from the DAG and run a breadth-first search~(BFS) on the subgraph. 
During the \texttt{TR\_CX} subgraph traversal, we check whether the boundary shapes match. 

If a boundary-shape mismatch is detected, we insert a \texttt{ROT} for the operand with the rotated boundary. 
In addition, we add two edges: one from the \texttt{ROT} to the unmatched \texttt{TR\_CX} and the other from the parent \texttt{TR\_CX} to the \texttt{ROT}. 
Lastly, we update the \texttt{ROT} target's boundary shape and propagate the change to the descendant \texttt{TR\_CX} instructions. 
In this manner, we reduce the total number of rotations by inserting \texttt{ROT} only when the boundary shapes are different. 

%%%
\subsubsection{Optimization \#2. Rotation scheduling in instruction scheduler} \label{subsubsec:4.3.3}
We exploit Opportunity \#2 by adding a rotation scheduling pass to the instruction scheduler.
We first schedule all logical instructions except \texttt{ROT} using the baseline algorithm (i.e., ALAP for \texttt{REQ} and ASAP for others). 
As a result, we obtain the earliest and latest possible timings for scheduling \texttt{ROT}, given the layer indices of the parent and child instructions.
Using this information, we can flexibly schedule \texttt{ROT} instructions. 
In this work, we propose two extreme rotation scheduling policies: aggregation and distribution. 

\textbf{Rotation aggregation.}
The aggregation policy aims to maximize the number of rotations within one layer. 
This policy is effective for the collective reflection-based rotation because the latency does not increase with the number of target LQs. 
In other words, the presence of patch rotations matters rather than their number. 
Therefore, the goal is to maximize the number of rotation-free instruction layers. 
We implement the aggregation by counting the number of rotations each layer can accommodate and greedily selecting the layer with the highest count. 

\textbf{Rotation distribution.}
In contrast, the distribution policy aims to minimize the number of rotations within one layer. 
We implement the distribution by sorting the rotation layer ranges and greedily selecting the smallest unused layer index. 
This policy serves as the worst-case scenario for reflection-based rotations, but it is necessary for an architectural idea whose disadvantage scales with the number of rotations per layer.

%%%
\subsubsection{Impact of compiler optimizations} \label{subsubsec:compopt_result}

\begin{table}[t]
\caption{Compiler optimization results}
\centering
\tabcolsep=2.0pt
\scriptsize
%\resizebox{\columnwidth}{!}{%
\begin{tabular}{|cc|c|c|c|c|c|c|c|c|c|}
\hline
\rowcolor[HTML]{AFABAB} 
\multicolumn{2}{|c|}{\cellcolor[HTML]{AFABAB}
  \begin{tabular}[c]{@{}c@{}}FTQC\\workload\end{tabular}} &
  \begin{tabular}[c]{@{}c@{}}Hei\end{tabular} &
  \begin{tabular}[c]{@{}c@{}}FH\end{tabular} &
  \begin{tabular}[c]{@{}c@{}}Jellium\end{tabular} &
  \begin{tabular}[c]{@{}c@{}}H4\end{tabular} &
  \begin{tabular}[c]{@{}c@{}}adder\end{tabular} &
  \begin{tabular}[c]{@{}c@{}}qft\end{tabular} &
  \begin{tabular}[c]{@{}c@{}}ising\\1d\end{tabular} &
  \begin{tabular}[c]{@{}c@{}}ising\\2d\end{tabular} &
  Average \\ \hline
\rowcolor[HTML]{FFFFFF} 
\multicolumn{1}{|c|}{\cellcolor[HTML]{FFFFFF}} &
  Baseline & 
  30.1	&
  20.5	&
  23.2	&
  25.5	&
  31.8	&
  68.7	&
  29.6	&
  57.8	&
  35.9  \\ \hhline{|~|-|-|-|-|-|-|-|-|-|-|}
\rowcolor[HTML]{FFFFFF} 
\multicolumn{1}{|c|}{\cellcolor[HTML]{FFFFFF}} &
  Skip &
  27.6 &	
  18.7 &	
  20.7 &	
  22.9 &
  31.8 &	
  68.7 &	
  29.5 &	
  57.7 &	
  34.7 \\ \hhline{|~|-|-|-|-|-|-|-|-|-|-|} 
\rowcolor[HTML]{FFFFFF} 
\multicolumn{1}{|c|}{\cellcolor[HTML]{FFFFFF}} &
  \begin{tabular}[c]{@{}c@{}}Skip+\\ Aggr.\end{tabular} &
  9.3  &	
  8.6  &	
  8.9  &	
  9.6  &
  21.2 &
  23.3 &
  20.4 &
  23.3 &
  15.6 \\ \hhline{|~|-|-|-|-|-|-|-|-|-|-|} 
\rowcolor[HTML]{FFFFFF} 
\multicolumn{1}{|c|}{\multirow{-5}{*}{\cellcolor[HTML]{FFFFFF}
\begin{tabular}[c]{@{}c@{}}
Layers\\with\\rotation\\(\%)\end{tabular}}} &
  \begin{tabular}[c]{@{}c@{}}Skip+\\ Dist.\end{tabular} &
  31.3 &
  19.8 & 
  21.2 & 
  24.0 & 
  37.1 & 
  90.2 & 
  37.1 &
  90.0 &
  43.8 \\ \hline
\rowcolor[HTML]{FFFFFF} 
\multicolumn{1}{|c|}{\cellcolor[HTML]{FFFFFF}} &
  \cellcolor[HTML]{FFFFFF}Baseline &
  100 &
  100 &
  100 &
  100 &
  100 &
  100 &
  100 &
  100 &
  100 \\ \hhline{|~|-|-|-|-|-|-|-|-|-|-|} 
\rowcolor[HTML]{FFFFFF} 
\multicolumn{1}{|c|}{\cellcolor[HTML]{FFFFFF}} &
  Skip &
  81.5 & 
  80.2 & 
  80.6 & 
  80.3 & 
  100  &
  100 & 
  99.8 & 
  99.9 & 
  89.8 \\ \hhline{|~|-|-|-|-|-|-|-|-|-|-|} 
\rowcolor[HTML]{FFFFFF} 
\multicolumn{1}{|c|}{\cellcolor[HTML]{FFFFFF}} &
  \begin{tabular}[c]{@{}c@{}}Skip+\\ Aggr.\end{tabular} &
  81.5 & 
  80.2 & 
  80.6 & 
  80.3 & 
  100  &
  100 & 
  99.8 & 
  99.9 & 
  89.8 \\ \hhline{|~|-|-|-|-|-|-|-|-|-|-|} 
\rowcolor[HTML]{FFFFFF} 
\multicolumn{1}{|c|}{\multirow{-5}{*}{\cellcolor[HTML]{FFFFFF}
\begin{tabular}[c]{@{}c@{}}
Rotation\\count\\(\%)\end{tabular}}} &
  \begin{tabular}[c]{@{}c@{}}Skip+\\ Dist.\end{tabular} &
  81.5 & 
  80.2 & 
  80.6 & 
  80.3 & 
  100  &
  100 & 
  99.8 & 
  99.9 & 
  89.8 \\ \hline
\rowcolor[HTML]{FFFFFF} 
\multicolumn{1}{|c|}{\cellcolor[HTML]{FFFFFF}} &
  \cellcolor[HTML]{FFFFFF}Baseline &
  \cellcolor[HTML]{FFFFFF}100 &
  100 &
  100 &
  100 &
  100 &
  100 &
  100 &
  100 &
  100 \\ \hhline{|~|-|-|-|-|-|-|-|-|-|-|}
\rowcolor[HTML]{FFFFFF} 
\multicolumn{1}{|c|}{\cellcolor[HTML]{FFFFFF}} &
  Skip &
  94.5& 
  94.9& 
  93.6& 
  94.2&
  100&
  100&
  99.9&
  100&
  97.1 \\ \hhline{|~|-|-|-|-|-|-|-|-|-|-|}
\rowcolor[HTML]{FFFFFF} 
\multicolumn{1}{|c|}{\cellcolor[HTML]{FFFFFF}} &
  \begin{tabular}[c]{@{}c@{}}Skip+\\ Aggr.\end{tabular} &
  54.3 &
  66.1 &
  61.6 &
  64.4 &
  82.5 &
  54.6 &
  81.5 &
  58.9 &
  64.7 \\ \hhline{|~|-|-|-|-|-|-|-|-|-|-|}
\rowcolor[HTML]{FFFFFF} 
\multicolumn{1}{|c|}{\multirow{-5}{*}{\cellcolor[HTML]{FFFFFF}
\begin{tabular}[c]{@{}c@{}}
Exec.\\time\\(\%)\end{tabular}}} &
  \begin{tabular}[c]{@{}c@{}}Skip+\\ Dist.\end{tabular} &
  102.0& 
  98.2&
  94.8&
  96.7&
  110.4&
  123.5&
  114.9&
  140.5&
  109.2 \\ \hline
\rowcolor[HTML]{FFFFFF} 
\multicolumn{1}{|c|}{\cellcolor[HTML]{FFFFFF}} &
  Baseline &
  11& 
  58&
  26&
  17&
  4 &
  9 &
  4 &
  7 &
  17\\ \hhline{|~|-|-|-|-|-|-|-|-|-|-|}
\rowcolor[HTML]{FFFFFF} 
\multicolumn{1}{|c|}{\cellcolor[HTML]{FFFFFF}} &
  Skip &
  11   & 
  58&
  26&
  17&
  4&
  8&
  4&
  7&
  16.9\\ \hhline{|~|-|-|-|-|-|-|-|-|-|-|}
\rowcolor[HTML]{FFFFFF} 
\multicolumn{1}{|c|}{\cellcolor[HTML]{FFFFFF}} &
  \begin{tabular}[c]{@{}c@{}}Skip+\\ Aggr.\end{tabular} &
  161 &  
  64&
  31&
  18&
  14&
  28&
  64&
  64&
  55.5 \\ \hhline{|~|-|-|-|-|-|-|-|-|-|-|}
\rowcolor[HTML]{FFFFFF} 
\multicolumn{1}{|c|}{\multirow{-5}{*}{\cellcolor[HTML]{FFFFFF}
\begin{tabular}[c]{@{}c@{}}
Max.\\per-layer\\rotations\end{tabular}}} &
  \begin{tabular}[c]{@{}c@{}}Skip+\\ Dist.\end{tabular} &
  1 &
  1 &
  1 &
  1 &
  3 &
  2 &
  2 &
  2 &
  1.6 \\ \hline
\end{tabular}%
%}
\label{tab:compopt}
\end{table}

\cref{tab:compopt} shows the impact of our compiler optimizations on the execution time and rotation counts. 
The Baseline without optimization suffers from severe rotation-time overhead, as 35.9\% of layers include rotation.
The rotation skip scheme (Skip) reduces the number of rotations by 10.2\% on average. 
However, this translates into only a 2.9\% reduction in execution time because most layers that originally contained rotations still include at least one rotation.
Building on this, by employing the aggregation policy (Skip+Aggr.), we achieve an average latency reduction of 35.3\%. 
We also show the flexibility of rotation scheduling via the distribution policy (Skip+Dist.), which reduces the number of per-layer rotations to one to three.

%%%%%%%%%%%%%%%%%%%%%%%%%%%%%%%%%%%%%%%%%%%%%%%%
\begin{figure}
	\centering
	\includegraphics[width=\figsize{}]{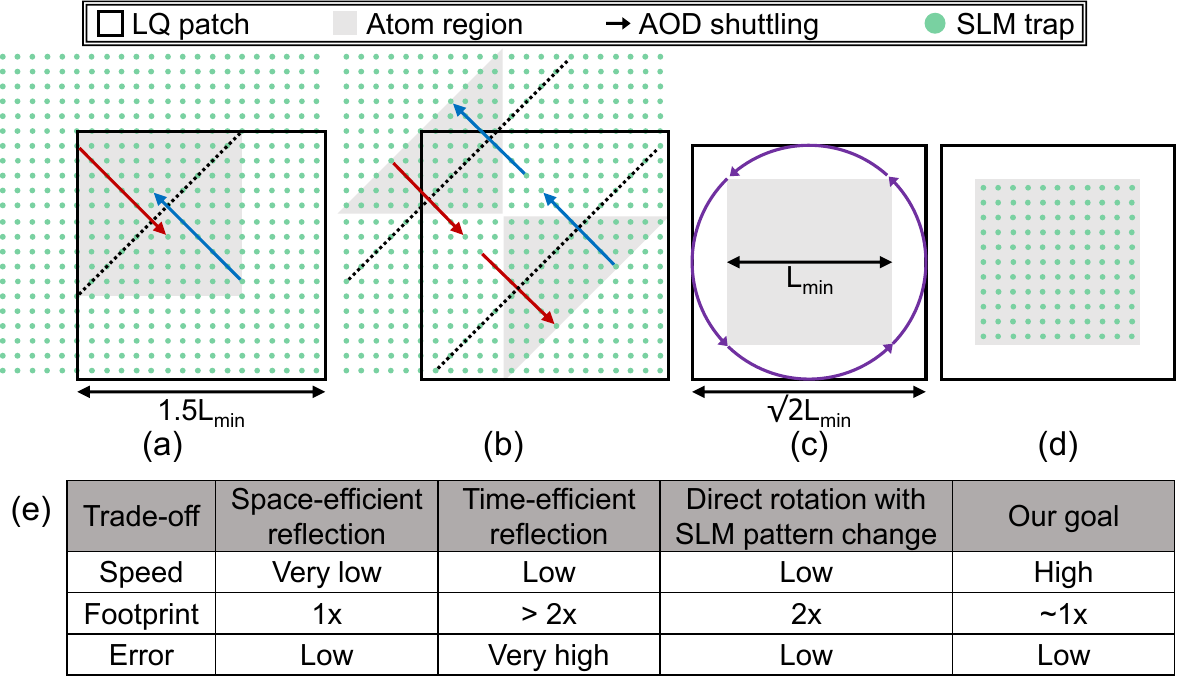}
	\caption{Limitations of conventional rotation methods: (a)--(b) time-efficient reflection requiring a larger footprint and incurring high error rates, and (c)--(d) direct rotation constrained by slow SLM pattern changes. (e) Summary of the rotation schemes and our goal.}
	\label{fig:fastrot_limit}
\end{figure}

\subsection{Full-stack architecture co-optimization} \label{subsec:4.3}
Even with our compiler optimizations, patch rotations remain the primary performance bottleneck due to the inherent latency of the space-efficient reflection protocol. 
Thus, we explore synergistic opportunities across the control hardware, qubit plane, and compiler stacks to eliminate the bottleneck.

%%%
\subsubsection{Motivation and limitations of prior art}
To accelerate rotations beyond the slow sequential swaps of the space-efficient reflection, we can consider time-efficient reflection (\cref{fig:fastrot_limit}(a)--(b)) or direct rotation (\cref{fig:fastrot_limit}(c)--(d)). 
However, naive implementations of these methods introduce prohibitive trade-offs. 
While the time-efficient reflection achieves speedups by parallelizing atom swaps, it demands at least a $2.25\times$ larger footprint and substantially increases the LER due to a significant increase in pick/drop operations (REFL\_TE in \cref{fig:fullstack_result}). 
Alternatively, while direct rotation is inherently fast, its execution requires the removal of static SLM traps to avoid collisions with moving atoms. 
Standard mitigation via runtime SLM pattern changes incurs millisecond-scale latencies that negate potential speedups (1~$\mathrm{ms}$ case, DIR\_CHANGE in \cref{fig:fullstack_result}).
In addition, we should double the footprint of all logical-qubit patches to support direct rotation at an arbitrary location.
We summarize the limitations of the aforementioned patch rotation methods in \cref{fig:fastrot_limit}(e). 

%%%
\subsubsection{\archname{} architecture}
To achieve high speed, low footprint, and low LERs simultaneously, we propose the \archname{} architecture through full-stack co-optimization with (1) direct rotation on (2) dedicated rotation patches, enabled by (3) distribution of rotations.

\begin{figure}
	\centering
	\includegraphics[width=\figsize{}]{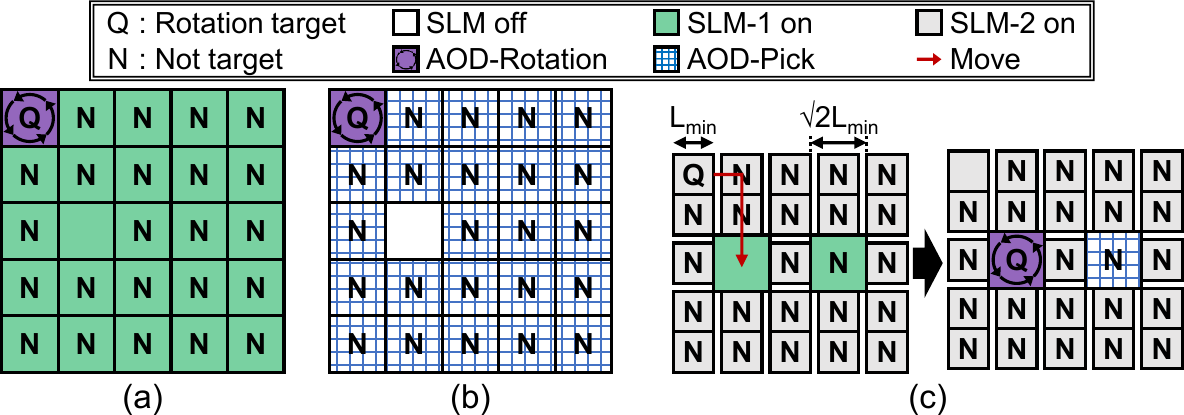}
	\caption{Proposed control hardware and qubit plane co-optimizations for direct rotation: (a)--(b) the SLM is temporarily turned off while non-target LQs are held by AODs during the rotation. (c) A heterogeneous qubit plane mitigates footprint overhead.}
	\label{fig:dirplane_opt}
\end{figure}

\begin{figure}
	\centering
	\includegraphics[width=\figsize{}]{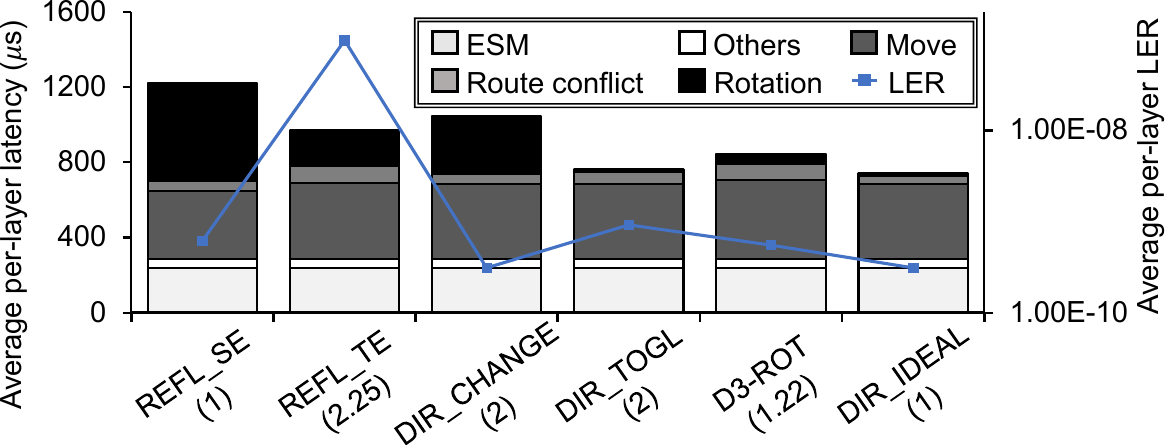}
	\caption{Average per-layer latency (bar), LER (line), and spatial footprint (indicated in labels) across various rotation implementations (averaged across FTQC workloads in \cref{tab:benchmark_programs}).}
	\label{fig:fullstack_result}
\end{figure}

\textbf{Control hardware: Direct rotation with SLM toggling.} 
Instead of slow SLM pattern changes, our key idea is to turn the SLM laser off (and on) to remove (and recover) traps with negligible latency (e.g., $1~\mu\mathrm{s}$~\cite{bluvstein2024logical, Graham2023multiscale}).
To prevent atoms from being lost from the traps, we pick all non-target atoms using separate AODs, safely turn off the SLM, rapidly perform the direct rotation on the target, and turn the SLM back on (\cref{fig:dirplane_opt}(a)--(b)).
As a result, we significantly reduce the average exposed rotation latency to $12.3~\mu\mathrm{s}$, corresponding to a $25.1\times$ reduction compared with SLM pattern change (DIR\_TOGL vs. DIR\_CHANGE in \cref{fig:fullstack_result}). 
However, this idea requires a $2\times$ footprint and incurs higher LERs due to increased pick/drop operations.

\textbf{Qubit plane: Dedicated rotation patches.}
To mitigate the footprint overhead, we additionally employ a heterogeneous qubit plane. 
We define two types of patches using separate SLM lasers: space-efficient normal patches (minimum footprint) and a few dedicated rotation patches (doubled footprint). 
As illustrated in \cref{fig:dirplane_opt}(c), rotation targets are shuttled to these dedicated patches for direct rotation. 
This restricts the footprint overhead to only the dedicated patches, keeping the rest of the plane compact.
We conservatively estimate the footprint overhead by scaling the width (height) of the columns (rows) containing the double-sized patches by a factor of $\sqrt{2}$.

\textbf{Compiler: Distribution of rotations.}
The heterogeneous qubit plane imposes a strict hardware constraint: the number of simultaneous rotations cannot exceed the number of dedicated rotation patches. 
We address this constraint by applying our rotation distribution compiler policy (\cref{subsubsec:4.3.3}). 
As previously discussed in \cref{subsubsec:compopt_result}, this policy elegantly bounds the per-layer rotation count to at most three for the evaluated workloads (\cref{tab:compopt}). 
Thus, three dedicated rotation patches are sufficient to meet the per-layer rotation demand of the evaluated workloads.
%%%
\subsubsection{Result summary}

\cref{fig:fullstack_result} summarizes the speed, footprint, and LER of our \archname{} architecture against intermediate implementations. 
We report the average per-layer latency in wall-clock time.
\archname{} successfully realizes a near-rotation-free architecture by reducing the average exposed rotation latency from $520~\mu\mathrm{s}$ (Baseline) to $49.8~\mu\mathrm{s}$. 
Notably, its performance gap compared to an ideal, zero-overhead direct rotation (DIR\_IDEAL) is a mere 13.8\% ($842.3~\mu\mathrm{s}$ vs. $740.1~\mu\mathrm{s}$ overall). 
By doubling the footprint of only three dedicated patches via rotation distribution, \archname{} limits the footprint overhead to just 22.4\% while maintaining a low LER.

%%%%%%%%%%%%%%%%%%%%%%%%%%%%%%%%%%%%%%%%%%%%%%%%
\subsection{D3-ROT: Ablation study} \label{subsec:ablation_study}

\begin{figure}
	\centering
	\includegraphics[width=\figsize{}]{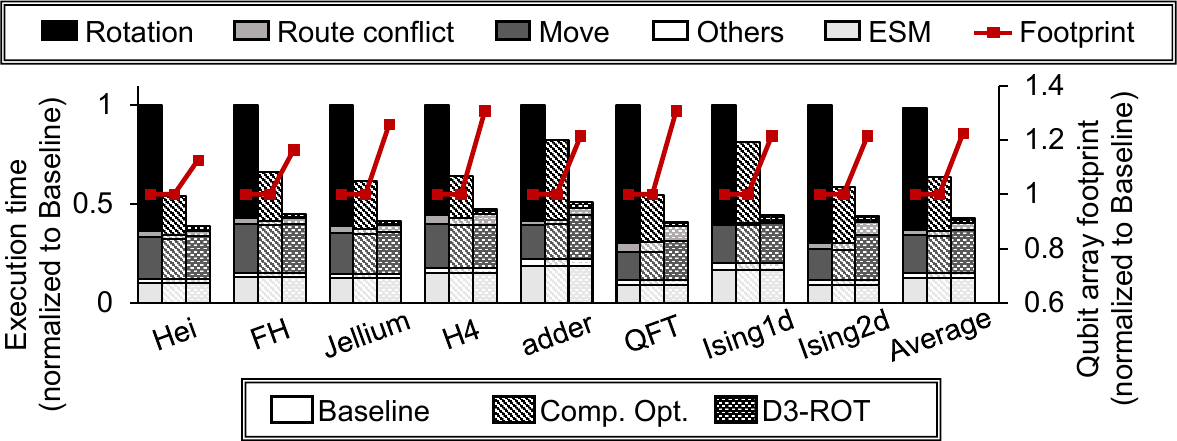}
	\caption{Ablation study of the proposed optimizations}
	\label{fig:opt_impact}
\end{figure}

To clarify the impact of our architecture exploration in \cref{sec:4}, we conduct an ablation study starting from the baseline architecture (Baseline).
We progressively apply our compiler optimization (Comp. Opt., \cref{subsec:compilier_opitimization}) and full-stack co-optimization (\archname, \cref{subsec:4.3}).
\cref{fig:opt_impact} illustrates the performance impact of these optimizations compared to the Baseline.
The left (right) y-axis represents the execution time (footprint) normalized to the Baseline. 

Across all FTQC workloads, the patch rotation dominates the execution time of Baseline, accounting for 61.9\% on average.
This severe execution bottleneck arises from the $O(d)$ latency of the naive reflection protocol in Baseline.
Therefore, we apply the rotation skip and aggregation techniques at the compiler level (Comp. Opt.).
As a result, we achieve a $1.55\times$ speedup, but the severe rotation bottleneck still remains (19.8--41.0\% in Comp. Opt.).

To resolve the bottleneck, our \archname{} architecture introduces the direct rotation protocol coupled with dedicated rotation patches and a rotation-distributing compilation strategy. 
As a result, our full-stack co-optimization nearly eliminates the patch rotation overhead and substantially reduces execution time.
In summary, our \archname{} achieves an average speedup of $2.27\times$ and up to $2.57\times$ (Hei), with a modest 22.4\% footprint overhead on average.

\section{Evaluation with lattice surgery architectures} \label{sec:evaluation}

%%%%%%%%%%%%%%%%%%%%%%%%%%%%%%%%%%%%%%%%%%%%%%%%
\subsection{Lattice surgery evaluation methodology} \label{subsec:experimental_setup}

For the lattice surgery (LS) baselines, we model architectures using Quration~\cite{suzuki2026quration}. 
We assume that each LS architecture consists of a 2D grid of surface-code patches, where each LQ is assigned to a $2 \times 2$ region, and the remaining patches are used temporarily for LS and code deformation, as in~\cite{beverland2022surface}. 
We translate each input quantum circuit into a sequence of LS instructions: (1) initializations to $\ket{0}$ or $\ket{+}$; (2) single- or two-qubit Pauli measurements in the X or Z basis; and (3) logical S and H gates implemented via code deformation. 
For T gates, we assume that a $\ket{\mathrm{T}}$ state is supplied by dedicated patches on the left side of the 2D grid and that the reaction time for gate teleportation equals the duration of $d$ ESM cycles. 
Two-qubit Pauli measurements and logical S and H gates require additional patches for LS and code deformation, and these patches are allocated using greedy algorithms.

We estimate the program success rate from the LER of a single patch over $d$ ESM rounds and the space-time volume (i.e., the integrated number of active surface-code patches over ESM rounds).
Consistent with the methodology described in \cref{subsubsec:success_probability}, we model the LER of a $d$-cycle memory operation as a function of $d$. 
The physical error parameters are the same as those in \cref{tab:eval_setup}. 
The total number of physical qubits is calculated by multiplying the number of surface-code patches by $2d^2-1$, the number of physical qubits per patch. 
The total execution time is obtained by multiplying the ESM duration by the total number of ESM rounds. 
For the ESM duration, we assume $237~\mu\mathrm{s}$ and $1~\mu\mathrm{s}$ for NA and superconducting platforms, respectively.

%%%%%%%%%%%%%%%%%%%%%%%%%%%%%%%%%%%%%%%%%%%%%%%%
\subsection{Comparison with lattice surgery}

\begin{figure}
	\centering
	\includegraphics[width=\figsize{}]{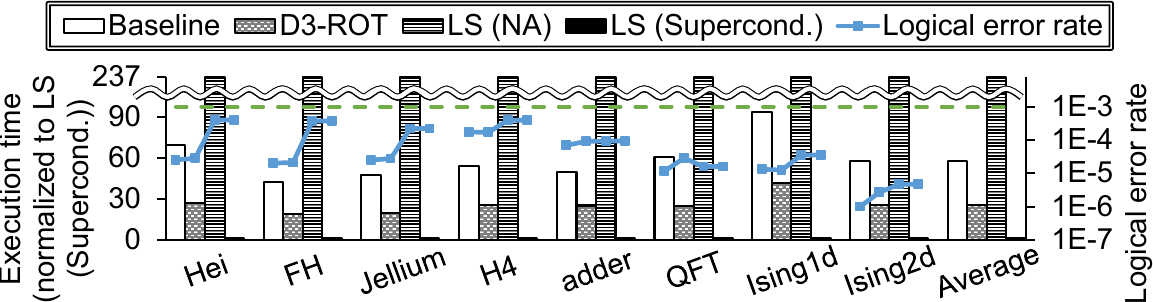}
	\caption{Comparison with NA-based and superconducting LS architectures.}
	\label{fig:vs_ls}
\end{figure}

\cref{fig:vs_ls} compares our \archname{} architecture with conventional LS architectures on NA and superconducting platforms.
The left y-axis represents the total execution time normalized to the superconducting LS baseline, while the right y-axis shows the overall program LER.
For each configuration, $d$ is selected to achieve a target program success rate of 99.9\% (\cref{tab:benchmark_programs}). 

A fundamental challenge for NA quantum computers is their intrinsically slow physical operations. 
Because the ESM latency on the NA platform ($237~\mu\mathrm{s}$) is substantially longer than that on the superconducting platform ($1~\mu\mathrm{s}$), the conventional NA-based LS architecture (LS (NA)) requires $237\times$ the execution time of its superconducting counterpart.

The baseline NA architecture (Baseline) mitigates this severe penalty by executing transversal logical gates in $O(1)$ ESM rounds.
However, despite this advantage, Baseline still requires $58\times$ the execution time of the superconducting LS architecture. 
By resolving the rotation bottleneck through full-stack co-optimization, our \archname{} architecture achieves a $2.27\times$ speedup over Baseline. 
Consequently, \archname{} reduces the execution time to $25.5\times$ that of the superconducting LS architecture.

We also analyze the space-time overhead for a more comprehensive comparison. 
LS inherently incurs substantial spatial overhead for routing and code deformation and requires a larger $d$ to achieve the same target program success rate (\cref{tab:benchmark_programs}). %%% space overhead geomean: 6.4198x
Despite the larger spatial overhead of LS, the baseline NA-based architecture still incurs $9.0\times$ space-time overhead relative to the superconducting LS baseline. 
In contrast, our \archname{} narrows this gap to $3.9\times$ by applying architectural solutions in a space-time-efficient manner.
This highlights the importance of architecture exploration tools and optimizations for NA-based FTQC systems.

%%%%%%%%%%%%%%%%%%%%%%%%%%%%%%%%%%%%%%%%%%%%%%%%
\subsection{Sensitivity analysis of AOD configurations} \label{subsec:sensitivity}

\begin{figure}
	\centering
	\includegraphics[width=\figsize{}]{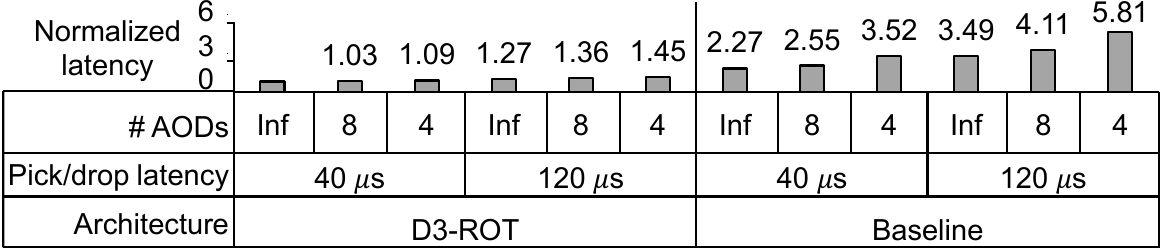}
	\caption{Sensitivity to AOD pick/drop latency and the number of AOD pairs.}
	\label{fig:sensitivity}
\end{figure}

We conduct a sensitivity analysis of various AOD setups.
\cref{fig:sensitivity} shows the execution times of Baseline and \archname{} for AOD configurations with as few as four AOD pairs, a conservative yet practical setting, and pick/drop latencies of up to $120~\mu\mathrm{s}$.

Reducing the number of AOD pairs to four incurs only a marginal 9\% slowdown.
By contrast, increasing the AOD pick/drop latency to 120~$\mu\mathrm{s}$ results in a 27\% slowdown, as pick/drop operations become the primary bottleneck in AOD-based atom shuttling.
Nevertheless, \archname{} shows only a 45\% slowdown even under these conservative AOD parameter settings.

In contrast, Baseline suffers from a significant $2.56\times$ slowdown under the same AOD setup.
This difference arises because the straightforward reflection-based patch rotation requires more AOD shuttling and pick/drop operations than the optimized rotation proposed in \archname{}.
As a result, the speedup of \archname{} over Baseline increases to $4.01\times$ under this conservative AOD configuration.

%%%%%%%%%%%%%%%%%%%%%%%%%%%%%%%%%%%%%%%%%%%%%%%%%%%%%%%%%%%%%%%%%%%%%%%%%%%%%%%%%%%%
\subsection{Resource-state distillation overhead}

\begin{table}[t]
\caption{Resource-state factory setup}
\centering
\tabcolsep=2.0pt
\scriptsize
%\resizebox{\columnwidth}{!}{%
\begin{tabular}{|c|ccc|cccc|c|}
\hline
\rowcolor[HTML]{AFABAB} 
 &
  \multicolumn{3}{c|}{\cellcolor[HTML]{AFABAB}\begin{tabular}[c]{@{}c@{}}Magic state\\cultivation~\cite{hirano2025efficient}\\(MSC) \end{tabular}} &
  \multicolumn{4}{c|}{\cellcolor[HTML]{AFABAB}\begin{tabular}[c]{@{}c@{}}Magic state\\distillation\\(MSD)\end{tabular}} &
  \begin{tabular}[c]{@{}c@{}}Y-state\\distillation\\(YSD)\end{tabular} \\ \hline
\rowcolor[HTML]{AFABAB} 
 &
  \multicolumn{1}{c|}{\cellcolor[HTML]{AFABAB}\begin{tabular}[c]{@{}c@{}}Color\\code\\(d=3)\end{tabular}} &
  \multicolumn{1}{c|}{\cellcolor[HTML]{AFABAB}\begin{tabular}[c]{@{}c@{}}Surface\\code\\(d=5)\end{tabular}} &
  \begin{tabular}[c]{@{}c@{}}Surface\\code\\(d=11)\end{tabular} &
  \multicolumn{2}{c|}{\cellcolor[HTML]{AFABAB}\begin{tabular}[c]{@{}c@{}}Lattice\\surgery\\ {\cite{litinski2019game}}\end{tabular}} &
  \multicolumn{2}{c|}{\cellcolor[HTML]{AFABAB}\begin{tabular}[c]{@{}c@{}}Transversal\\ {\cite{sunami2025transversalsurfacecodegamepowered}}\end{tabular}} &
  \begin{tabular}[c]{@{}c@{}}Transversal\\ {\cite{sunami2025transversalsurfacecodegamepowered}}\end{tabular} \\ \hline
\# patches &
  \multicolumn{1}{c|}{4} &
  \multicolumn{1}{c|}{4} &
  1 &
  \multicolumn{2}{c|}{11} &
  \multicolumn{2}{c|}{7} &
  2 \\ \hline
  \begin{tabular}[c]{@{}c@{}}
  Time cost\end{tabular} &
  \multicolumn{1}{c|}{$4\tau{}_{\mathrm{esm}}$} &
  \multicolumn{1}{c|}{$3\tau{}_{\mathrm{esm}}$} &
  $15\tau{}_{\mathrm{esm}}$ &
  \multicolumn{2}{c|}{$11d\tau{}_{\mathrm{esm}}$} &
  \multicolumn{2}{c|}{
  \begin{tabular}[c]{@{}l@{}}
  $17\tau{}_{\mathrm{mov}}$\\$+18\tau{}_{\mathrm{esm}}$\end{tabular}} &
  \begin{tabular}[c]{@{}l@{}}
  $2\tau{}_{\mathrm{mov}}$\\$+\tau{}_{\mathrm{rot}}$\\$+4\tau{}_{\mathrm{esm}}$\end{tabular} \\ \hline
 &
  \multicolumn{1}{c|}{\begin{tabular}[c]{@{}c@{}}Survival\\rate\end{tabular}} &
  \multicolumn{1}{c|}{\begin{tabular}[c]{@{}c@{}}Survival\\rate\end{tabular}} &
  \begin{tabular}[c]{@{}c@{}}Survival\\rate\end{tabular} &
  \multicolumn{1}{c|}{\#$ \ket{\mathrm{T}}$} &
  \multicolumn{1}{c|}{\#$ \ket{\mathrm{Y}}$} &
  \multicolumn{1}{c|}{\#$ \ket{\mathrm{T}}$} &
  \multicolumn{1}{c|}{\#$ \ket{\mathrm{Y}}$} &
   \\ \cline{2-9} 
\multirow{-3}{*}{Note} &
  \multicolumn{1}{c|}{0.822} &
  \multicolumn{1}{c|}{0.770} &
  0.568 &
  \multicolumn{1}{c|}{15} &
  \multicolumn{1}{c|}{0} &
  \multicolumn{1}{c|}{15} &
  15 &
   \\ \hline
\end{tabular}%
%}
\label{tab:factory}
\end{table}

\begin{figure}
	\centering
	\includegraphics[width=\figsize{}]{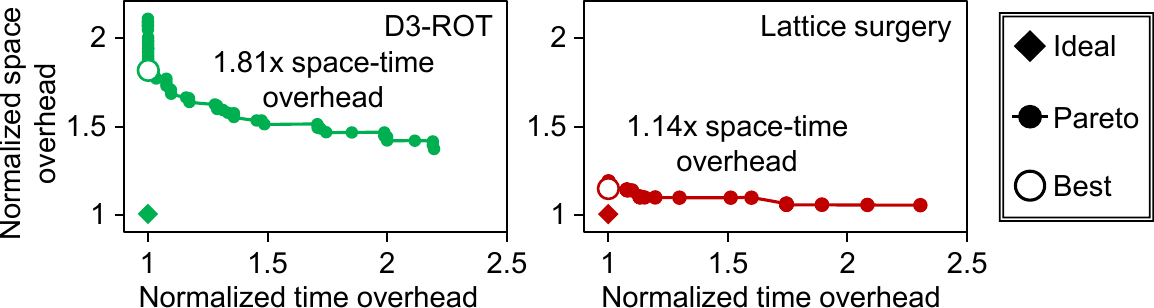}
	\caption{Resource-state distillation overhead.}
	\label{fig:distillation}
\end{figure}

We evaluate the resource-state (e.g., $\ket{\mathrm{T}}$, $\ket{\mathrm{Y}}$) distillation overhead for the QPE workloads. 
\cref{tab:factory} summarizes the assumed factory setup, where $\tau{}_{\mathrm{esm}}$, $\tau{}_{\mathrm{rot}}$, and $\tau{}_{\mathrm{mov}}$ indicate the latencies of an ESM round, patch rotation, and single-patch move for transversal CX.

%%%
\subsubsection{Resource-state factory cost}
\cref{tab:benchmark_programs} shows that the large QPE workloads contain more than $10^{9}$ gates, requiring a target LER below $10^{-12}$. 
To meet this target LER, we adopt a hybrid approach in which magic state cultivation~(MSC) produces the input $\ket{\mathrm{T}}$ states for subsequent magic state distillation~(MSD).

\textbf{Magic state distillation (MSD).}
For MSD, we use the 15-to-1 protocol, which is widely adopted in recent FTQC studies. 
For the LS and transversal surface-code architectures, we use the efficient implementations proposed in~\cite{litinski2019game} and~\cite{sunami2025transversalsurfacecodegamepowered}, respectively.

\textbf{Magic state cultivation (MSC).}
Because the 15-to-1 MSD suppresses an input error rate $p$ to approximately $35p^{3}$, an MSC output error rate on the order of $10^{-5}$ is sufficient. 
Therefore, we adopt an efficient MSC protocol~\cite{hirano2025efficient}, which can produce a clean-boundary patch with an output error rate of $3\times10^{-6}$ at a physical error rate of $10^{-3}$.

\textbf{Y-state distillation (YSD).}
Our S gate and MSD implementations also require distilled $\ket{\mathrm{Y}}$ states.
Therefore, we include the Y-state distillation overhead, following the implementation in~\cite{sunami2025transversalsurfacecodegamepowered}.

%%%
\subsubsection{Dynamic trace-based simulation}
For each resource-state-factory configuration, defined by the number and cost of factories, we run a dynamic trace-based simulation. 
Specifically, for each statically scheduled instruction layer, we check whether a sufficient number of resource states are available. 
If not, we insert a delay until the required resource states become available through probabilistic factory executions.
The aggregated delay represents the performance overhead caused by insufficient resource-state production throughput.

%%%
\subsubsection{Space-time overhead analysis}
\cref{fig:distillation} shows the space-time overhead of resource-state factories for \archname{} and the LS architectures. 
For a comprehensive analysis, we sweep the number of factories of each type and report the Pareto frontiers showing the trade-offs between time and space overheads. 
In summary, resource-state factories increase the space-time costs of \archname{} and the LS architectures by factors of $1.81$ and $1.14$, respectively. 
We attribute the higher space-time overhead of \archname{} mainly to two factors. 
First, MSC is more costly in transversal architectures because its throughput is low relative to the rate at which logical operations are executed. 
Second, MSD in \archname{} requires 15 distilled $\ket{\mathrm{Y}}$ states, directly adding to the overhead. 
Nevertheless, this $1.81\times$ space-time cost multiplier remains manageable relative to the $61\times$ reduction in space-time cost compared with the NA-based LS architecture.

\section{Discussion}

\begin{figure}
	\centering
	\includegraphics[width=\figsize{}]{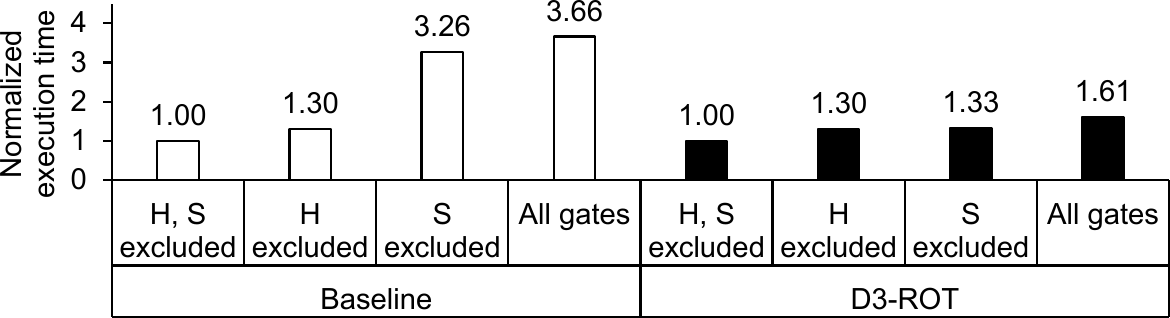}
	\caption{Breakdown of the contribution of the H and S gates to the total execution time.}
	\label{fig:hs_breakdown}
\end{figure}

\subsection{Comparison with prior resource analysis}
We compare our work with a prior resource analysis study on NA-based FTQC systems~\cite{Zhou2025resource}. 
The previous analysis provides first-order estimates for a few hand-crafted circuits, but does not fully capture the detailed architectural bottlenecks studied in this paper. 
In contrast, our work (1) provides a tool for analyzing previously overlooked architecture-level bottlenecks, (2) demonstrates architectural optimization as a case study, and (3) enables a more in-depth space-time analysis beyond first-order estimates.

Most notably, the previous analysis assumes equal atom-shuttling times for transversal H and S gates and transversal entangling gates, which can lead to inaccurate resource estimates.
To illustrate the importance of carefully considering logical H and S gates, \cref{fig:hs_breakdown} quantifies their contribution to the total execution time.

In Baseline, H and S gates account for as much as 73\% of the total latency, highlighting the importance of accurately modeling their overhead and motivating our architectural design.
Furthermore, even in the optimized \archname{}, H and S gates account for up to 38\% of the total latency, indicating that they remain significant contributors to execution time.
We also emphasize that, in the folded-transversal implementation assumed in~\cite{Zhou2025resource}, the S gate incurs an overhead comparable to that of the H gate in Baseline, as both rely on high-latency diagonal reflections.

\subsection{Relevance to qLDPC codes}
Our work focuses on the transversal surface code as the first target for architecture-level modeling and optimization. 
This section discusses the relevance of our work to architectures employing quantum low-density parity-check (qLDPC) codes.

First, our architectural insights and optimizations remain valid in load-store architectures with separate compute and memory subsystems.
Recent studies on such architectures commonly use surface codes for fast computation and qLDPC codes for high-density memory storage~\cite{qsieve}. 
Therefore, our simulation framework and architectural designs developed for surface-code-based systems can also contribute to architectures that employ qLDPC codes.
Second, our full-stack modeling methodology is broadly applicable to qLDPC codes.
We design NAQsim by carefully considering common modules across different QEC codes, such as the instruction translator, scheduler, and mapper in the compilation stage, as well as micro-operation decomposition and scheduling in the execution stage.
Therefore, our simulator can be extended to qLDPC codes through appropriate modifications to the corresponding modules.

\subsection{Relation to prior neutral-atom compilers}
Despite extensive prior work on neutral-atom compilers, these techniques cannot be directly applied to FTQC architectures.
First, several prior studies assume hardware models that differ substantially from large-scale FTQC (e.g., fixed-array architectures without atom shuttling~\cite{geyser}).
Second, even state-of-the-art zoned-architecture compilers (e.g., PowerMove~\cite{powermove} and ZAC~\cite{zac}) do not fully account for key FTQC constraints.
For example, they do not account for the overhead of fault-tolerant single-qubit gates (e.g., T, H, and S) and the FTQC-specific constraints of AOD shuttling (e.g., restricted movement paths and route conflicts in \cref{subsubsec:3.3.2}).

Nevertheless, many ideas developed in prior neutral-atom compilers remain valuable for FTQC, including logical-qubit mapping, CX-gate placement, and AOD grouping strategies~\cite{powermove, zac}. 
Accordingly, we adopt these techniques from ZAC~\cite{zac} as our baseline and extend them to support FTQC architecture simulation.
We emphasize that our primary architectural contribution, identifying and mitigating the patch rotation bottleneck in NA-based FTQC systems, is orthogonal to existing compiler optimizations. 
As such, our methodology can naturally incorporate prior compiler techniques whenever applicable in FTQC systems.

\section{Related work}

\textbf{FTQC simulation framework.}
Byun et al.~\cite{xqsim} and Min et al.~\cite{qisim} proposed FTQC simulation frameworks for superconducting quantum computers using lattice surgery. 
Quration~\cite{suzuki2026quration} is a quantum resource estimation toolchain that supports lattice-surgery-based FTQC across different qubit technologies. 
However, no existing FTQC modeling tool can address the unique capabilities, control hardware constraints, and qubit plane features of the neutral-atom platform, which can employ more efficient FTQC protocols.

\textbf{Resource analysis for neutral-atom FTQC.}
Zhou et al.~\cite{Zhou2025resource} conducted a resource analysis for large-scale FTQC algorithms with neutral atoms. 
However, the prior work ignored many logical operations, control-hardware constraints, and limitations on the qubit-plane footprint, leaving the analysis at a high level. 

To the best of our knowledge, our work is the first study to develop a modeling tool for FTQC on a neutral-atom platform and conduct detailed architecture-level analysis and optimization to address the performance bottleneck.

\section{Conclusion}
In this work, we proposed NAQsim, an open-source simulation framework for exploring various FTQC architectures for neutral-atom quantum computers. 
Using NAQsim, we further proposed a near-rotation-free architecture (\archname{}) through full-stack co-optimization and addressed the previously overlooked performance bottleneck in the neutral-atom platform.
Our evaluation using practical FTQC benchmarks showed that \archname{} achieves a significant speedup, underscoring the importance of NAQsim for developing fast, space-efficient FTQC with neutral atoms.

\clearpage
\section*{Appendix: Artifact Evaluation}

%%%%%%%%%%%%%%%%%%%%%%%%%%%%%%%%%%%%%%%%%%%%%%%%%%%%%%%%%%%%%%%%%%%%%
\subsection{Abstract}
This artifact provides the Zenodo-hosted reproducibility package for the NAQsim framework and the experimental workflow used in the paper.
It includes the benchmark datasets and figure-generation scripts.
It reproduces the numerical results shown in \cref{fig:logical_error_model}(a)--(b), \cref{fig:rot_bottleneck}(a), Table~\ref{tab:compopt}, and
Figs.~\ref{fig:fullstack_result}--\ref{fig:hs_breakdown}.
Two evaluation modes are supported: (i) a fast path that regenerates the CSV tables and Matplotlib figures from the archived raw data, and (ii) an end-to-end path that reruns the compilation and architecture execution stages.
Figs.~\ref{fig:vs_ls} and \ref{fig:distillation} additionally use the external Quration~\cite{suzuki2026quration} \texttt{qret} executable.
No GPU, quantum processor, or other specialized hardware is required.
NAQsim is released as open-source software at \url{https://github.com/naqsim/naqsim}.
The artifact version evaluated by the MICRO 2026 Artifact Evaluation Committee is archived at \url{https://zenodo.org/records/21542759}. 
A subsequent version adding the omitted sensitivity-analysis raw data without changing the code, evaluation procedure, or reported results, is available at \url{https://doi.org/10.5281/zenodo.22684078}.
Detailed build and execution instructions are provided in the artifact \path{README.md}.

\subsection{Artifact check-list (meta-information)}

{\small
\begin{itemize}
  \item {\bf Algorithm: } Dependency-aware logical-qubit scheduling, restricted AOD routing, compiler optimizations, and magic-state resource sweeps.
  \item {\bf Program: } NAQsim (Python), Bash entry points, and Quration~\cite{suzuki2026quration} \texttt{qret} (C++ executable; required for Figs.~\ref{fig:vs_ls} and \ref{fig:distillation}).
  \item {\bf Compilation: } Quantum-circuit preprocessing and schedule construction in Python~3; the tested Python environment is managed with \texttt{uv}~0.8.11. Quration is built from source using CMake and \texttt{vcpkg}.
  \item {\bf Binary: } External \texttt{qret}; its path is supplied with \texttt{--quration-bin}. All other artifact components are Python or shell scripts and require no separate compilation.
  \item {\bf Data set: } QASM benchmark circuits (the paper's QPE benchmark set and FTCircuitBench~\cite{ftcircuitbench}).
  \item {\bf Run-time environment: } A Linux or macOS shell environment with Python~3. The fast path was also tested under WSL2, and the end-to-end simulations were tested on Linux servers.
  \item {\bf Hardware: } CPU only. End-to-end simulation is CPU- and memory-intensive.
  \item {\bf Run-time state: } End-to-end runs maintain a shared, content-addressed cache under \path{artifact_evaluation/.cache/} and store simulator outputs under \path{src/tsc_naa_sim/output/}. Cached compilation and execution results are reused across figures and tables.
  \item {\bf Execution: } Three Bash wrappers invoke Python through \texttt{uv}.
  \path{run_replotting.sh} regenerates all plots used in the paper from the archived CSV files.
  \path{run_full_simulation.sh} runs the compilation producers, execution producers, and analysis stages; with no workload argument, it then invokes \path{run_full_rendering.sh}. An optional workload argument restricts the simulation to one benchmark. A Slurm array wrapper assigns one workload to each job, and a dependent rendering job invokes \path{run_full_rendering.sh} after all workloads finish.
  \item {\bf Metrics: } Scheduled layer latency and execution time (in microseconds), latency breakdown by logical operation and routing/rotation conflict, normalized speedup, logical-qubit-plane footprint, LER, and magic-state distillation space, time, and space-time cost.
  \item {\bf Output: } CSV files and PDF/PNG figures under each target's \path{results/} directory.
  \item {\bf Experiments: } \cref{fig:logical_error_model}(a)--(b), \cref{fig:rot_bottleneck}(a), Table~\ref{tab:compopt}, and Figs.~\ref{fig:fullstack_result}--\ref{fig:hs_breakdown}.
  \item {\bf How much disk space required (approximately)?: } 30\,GB.
  \item {\bf How much time is needed to prepare workflow (approximately)?: } Less than 30 minutes for the Quration build and environment setup.
  \item {\bf How much time is needed to complete experiments (approximately)?: } Replotting all supplied data takes one minute on a laptop-class CPU. The full experiments require approximately 30 hours on a server with an AMD EPYC 9684X processor.
  \item {\bf Publicly available?: } Yes.
  \item {\bf Code licenses (if publicly available)?: } MIT.
  \item {\bf Data licenses (if publicly available)?: } MIT.
  \item {\bf Workflow automation framework used?: } We prepared three Bash wrappers ( \path{run_replotting.sh}, \path{run_full_simulation.sh}, and \path{run_full_rendering.sh}) and prepared Slurm batch job scripts for the end-to-end simulation.
  \item {\bf Archived (provide DOI)?: } 10.5281/zenodo.21542759
\end{itemize}
}

%%%%%%%%%%%%%%%%%%%%%%%%%%%%%%%%%%%%%%%%%%%%%%%%%%%%%%%%%%%%%%%%%%%%%
\subsection{Description}

\subsubsection{How to access}
The artifact-evaluation package is available at \url{https://zenodo.org/records/21542759}.
After extraction, the AE entry points are located in \path{NAQsim/src/tsc_naa_sim/artifact_evaluation/}.
The accompanying \path{README.md} describes the experiment workflow.

\subsubsection{Hardware dependencies}
No accelerator or quantum hardware is needed.
Raw-data analysis is lightweight and was tested on a laptop-class x86-64 CPU.
End-to-end simulations are CPU- and memory-intensive.
We recommend a high-memory Linux server with at least 32 threads and 30\,GB of free disk space.
We tested the end-to-end workflow on a server with an AMD EPYC 9684X processor and 768\,GB of memory; it took approximately 30 hours.

\subsubsection{Software dependencies}
The tested environment uses Python 3.13.3 and \texttt{uv}~0.8.11.
Python dependencies are declared in \path{pyproject.toml}.
The tested Quration~\cite{suzuki2026quration} commit is \texttt{293912c18ee6}.

\subsubsection{Data sets}
Our benchmark suite contains QASM files for four qubitization-based QPE workloads (2D Heisenberg, Fermi--Hubbard, Jellium, and H4) and four circuits from FTCircuitBench~\cite{ftcircuitbench} (a quantum adder, QFT, and quantum simulations of 1D and 2D Ising models based on Trotterization).
The workload manifest is \path{artifact_evaluation/config/workloads.csv}.
It also records the code distances listed in Table~\ref{tab:benchmark_programs}, along with the repetition counts and default worker counts.

The archived numerical inputs under \path{artifact_evaluation/raw_data/} are organized by the corresponding paper target and the workflow stage that produced them.
\cref{fig:logical_error_model} uses two CSV files generated at high physical error rates for held-out $d=23$ validation and two low-error CSV files from the accepted version for model fitting.
The other directories contain compiler, D3-ROT, final-analysis, sensitivity, Quration, magic-state, and H/S-skipping CSVs.
\path{raw_data/README.md} gives the exact schema and missing-input audit.
The raw-data tree is treated as read-only input: analyzers write to a separate \path{<target>/results/} directory.

%%%%%%%%%%%%%%%%%%%%%%%%%%%%%%%%%%%%%%%%%%%%%%%%%%%%%%%%%%%%%%%%%%%%%
\subsection{Installation}
Install \texttt{uv} and create the locked Python environment.

{
\footnotesize
\begin{verbatim}
cd /path/to/NAQsim/ && uv sync
\end{verbatim}
}

\noindent
For Figs.~\ref{fig:vs_ls} and \ref{fig:distillation}, clone and build Quration according to its \path{README.md}.
Note the absolute path to the \texttt{qret} binary.

%%%%%%%%%%%%%%%%%%%%%%%%%%%%%%%%%%%%%%%%%%%%%%%%%%%%%%%%%%%%%%%%%%%%%
\subsection{Experiment workflow}
Run all commands below from \path{NAQsim/src/tsc_naa_sim/artifact_evaluation/}.

\subsubsection{Fast validation from the archived data}
This is the recommended first test and takes less than one minute on a laptop-class CPU.
It does not run the architecture simulator or Quration.
It recursively reads the CSV files under \path{raw_data/} and writes new results under each target's \path{results/} directory:

\noindent
{
\footnotesize
\begin{verbatim}
./run_replotting.sh
\end{verbatim}
}

\noindent
The command renders \cref{fig:logical_error_model}(a)--(b), \cref{fig:rot_bottleneck}(a), Table~\ref{tab:compopt}, and Figs.~\ref{fig:fullstack_result}--\ref{fig:hs_breakdown}.

\subsubsection{Full simulation}
A complete end-to-end full simulation consists of the Quration prerequisite followed by the architecture-simulation workflow.
Generate the Quration results before invoking \path{run_full_simulation.sh}.
After building Quration, pass the absolute path to \texttt{qret}:

\noindent
{
\footnotesize
\begin{verbatim}
./quration/run.sh \
    --quration-bin /path/to/quration/build/main/qret
\end{verbatim}
}

\noindent
The default output root is \path{quration/results/}.
If \texttt{--output-dir DIR} is used, pass the same directory to the simulator via \texttt{--quration-results-dir DIR}.

After the prerequisite completes, the following command selects every workload from \path{config/workloads.csv}, populates the cache with all unique compilation results, executes the simulator, performs the target-specific analyses, and finally invokes \path{run_full_rendering.sh} to render the paper-facing figures and table:

{
\footnotesize
\begin{verbatim}
./run_full_simulation.sh
\end{verbatim}
}

The reported evaluation results use the simple row-major order mapping selected by \texttt{--use-naive-mapping}, while NAQsim also supports the SA-based LQ mapper described in \cref{subsubsec:logical_qubit_mapper}.
This execution takes approximately 30 hours on the tested AMD EPYC 9684X server.
We recommend one of the optional workflows in Sections E3 and E4.

\subsubsection{(Optional) Workload-wise execution\label{subsubsec:workload-wise}}
H4 is the lightest QPE workload and provides a relatively quick check of the full-simulation workflow.
First, generate the corresponding Quration results for H4, and then run all applicable simulation targets
with four workers:

{
\footnotesize
\begin{verbatim}
./quration/run.sh --workload h4 \
    --quration-bin /path/to/quration/build/main/qret
./run_full_simulation.sh h4 --num-threads 4
\end{verbatim}
}

For debugging or separate batch allocations, run a target one stage at a time:

{
\footnotesize
\begin{verbatim}
./fig11a/run.sh --workload h4 --stage compilation
./fig11a/run.sh --workload h4 --stage execution
./fig11a/run.sh --workload h4 --stage analysis
uv run python fig11a/analyze.py --workload h4
\end{verbatim}
}

\noindent
To select another target, replace \path{fig11a} with \path{table3}, \path{fig15}, \path{fig16}, \path{fig17}, \path{fig18}, \path{fig19}, or \path{fig20}. 

For a complete workload-wise reproduction, first generate the full Quration prerequisite without \texttt{--workload}, as described above.
Then run \path{run_full_simulation.sh} once for each of the eight workloads.
After the workload-wise simulation and analysis runs for all eight workloads have completed, render the complete figures and table once:

{
\footnotesize
\begin{verbatim}
./run_full_rendering.sh
\end{verbatim}

}

\subsubsection{(Optional) Execution on a Slurm cluster\label{subsubsec:slurm}}
After generating the complete Quration prerequisite, create the log directory, submit the supplied workload array, and make the rendering job depend on the completion of the array:

{
\footnotesize
\begin{verbatim}
mkdir -p logs
SIM_JOB_ID=$(sbatch \
    --parsable batch_run_full_simulation.sh)
sbatch --dependency=afterok:"${SIM_JOB_ID%%;*}" \
    batch_run_full_rendering.sh
\end{verbatim}
}

The Slurm wrappers are cluster-specific.
Users must adapt the partition and wall-time settings and virtual-environment activation commands to their environment.
The full experiment requires around 30 hours on the tested AMD EPYC 9684X server.

%%%%%%%%%%%%%%%%%%%%%%%%%%%%%%%%%%%%%%%%%%%%%%%%%%%%%%%%%%%%%%%%%%%%%
\subsection{Evaluation and expected results}
Successful analyzer runs create CSV, PDF, and PNG files.
The principal outputs are:
\begin{itemize}
  \item \cref{fig:logical_error_model}: \path{fig10a.*}, \path{fig10b.*}, CSVs containing the plotted data, and \path{logical_error_model_coefficients.csv}.
  \item \cref{fig:rot_bottleneck}(a) and Table~\ref{tab:compopt}: a normalized layer-latency breakdown and results for the four compiler configurations (Baseline, Skip, Skip+Aggr., and Skip+Dist.).
  \item \cref{fig:fullstack_result}: latency breakdown and LER for the six routing/rotation configurations, including D3-ROT.
  \item \cref{fig:opt_impact}: execution-time breakdown and normalized qubit-plane footprint
    for Baseline, compiler optimization, and D3-ROT.
  \item \cref{fig:vs_ls}: normalized execution time and LER for NAQsim and the Quration-derived lattice-surgery baselines.
  \item \cref{fig:sensitivity}: the sensitivity of geometric-mean latency to AOD pick/drop time and the number of AODs
  \item \cref{fig:distillation}: D3-ROT and lattice-surgery magic-state space-time tradeoffs and their Pareto frontiers.
  \item \cref{fig:hs_breakdown}: normalized execution time and its breakdown when H/S-gate skipping is enabled.
\end{itemize}

%%%%%%%%%%%%%%%%%%%%%%%%%%%%%%%%%%%%%%%%%%%%%%%%%%%%%%%%%%%%%%%%%%%%%
\subsection{Methodology}
Submission, reviewing and badging methodology:
\begin{itemize}
  \item \url{https://www.acm.org/publications/policies/artifact-review-and-badging-current}
  \item \url{https://cTuning.org/ae}
\end{itemize}

%%%%%
\bibliographystyle{./IEEEtran}
\bibliography{refs}

\end{document}